\documentclass[twocolumn,english,superscriptaddress]{revtex4-2}
\usepackage[T1]{fontenc}
\usepackage[latin9]{inputenc}

\usepackage{varwidth}

\usepackage{verbatim}
\usepackage{amsmath}
\usepackage{amssymb}
\usepackage{graphicx}
\usepackage{color}
\usepackage[dvipsnames]{xcolor}

\providecommand{\tabularnewline}{\\}
\makeatletter
\usepackage[dvips,letterpaper,margin=0.75in,bottom=0.5in]{geometry}

\providecommand{\tabularnewline}{\\}
\newenvironment{cellvarwidth}[1][t]
    {\begin{varwidth}[#1]{\linewidth}}
    {\@finalstrut\@arstrutbox\end{varwidth}}

\makeatother

\usepackage{babel}

\newcommand{\tj}[1]{{\color{orange} #1}}

\begin{document}
\title{Symmetry as a route to generalized bosonic Kitaev chains}

\author{Gideon Lee} 
\affiliation{Pritzker School of Molecular Engineering, The University of Chicago,
Chicago, Illinois 60637, USA}
\author{Tony Jin} 
\affiliation{Universit\'e C\^ote d'Azur, CNRS, Centrale Med, Institut de Physique de Nice, 06200 Nice, France}
\author{Aashish A. Clerk}
\affiliation{Pritzker School of Molecular Engineering, The University of Chicago,
Chicago, Illinois 60637, USA}

\begin{abstract}
The bosonic Kitaev chain (BKC) model is a deceptively simple looking quadratic pairing Hamiltonian.  Despite being purely Hermitian, it exhibits a number of striking non-Hermitian topological phenomena, including skin effects.  We show here how symmetries play a key role in this model, and how identifying these allows one to develop  generalized BKC-like models.  We emphasize the surprising fact that any quadratic bosonic pairing Hamiltonian with a sublattice (chiral) symmetry {\it necessarily} has a dynamical matrix with an effective time reversal symmetry.  This symmetry is unrelated to physical time-reversal, but enables non-trivial topological invariants.  We also discuss how this symmetry is unrelated to another key property of the BKC, the decoupling of quadrature dynamics.  This feature can instead be
connected to a distinct symmetry, namely an effective particle-hole symmetry of the dynamical matrix.  We discuss non-trivial generalized BKC models that only keep one of these two effective symmetries intact.  We also provide a classification of all translationally-invariant 1D pairing Hamiltonians, and show connections between the BKC and a well-studied non-Hermitian fermionic system, the symplectic Hatano-Nelson model.    
\end{abstract}
\maketitle







\section{Introduction}


The bosonic Kitaev chain (BKC) model is an innocuous looking quadratic model of bosons in 1D, with both hopping and pairing terms on each nearest-neighbor bond.  Its form mimics the well-known fermionic Kitaev Majorana chain.  The basic BKC Hamiltonian is
\cite{McDonald_PRX_2018_BKC} : 
\begin{equation}
\label{eq:H.BKC.OG}
\begin{aligned}\hat{H}_{\rm BKC} & = \frac{1}{2}\sum_{j}\left(iw\hat{a}_{j+1}^{\dag}\hat{a}_{j}+i\Delta\hat{a}_{j+1}^{\dag}\hat{a}_{j}^{\dag}+\text{H.c}.\right),\end{aligned}
\end{equation}
where
$\hat{a}_{j}$ are \textit{bosonic} operators, $[\hat{a}_{i},\hat{a}_{j}^{\dag}]=\delta_{ij}$, and $w$, $\Delta$ are real parameters of the model. 

The behaviour of this simple model is extremely rich. Despite being described by a Hermitian Hamiltonian, it exhibits striking features that are typically viewed as hallmarks of non-Hermitian physics (see Refs.~\cite{Okuma_2022_NH_review, Bergholtz_2021_RMP_NH_review, Ashida_2020_NH_review, Lin_2023_frontiers_topological_NH_review} for recent reviews).  There are two key relevant features:
\begin{enumerate}
    \item The Hermitian BKC model manifests the non-Hermitian skin effect (NHSE):  for periodic boundary conditions (PBC) all eigenstates are delocalized planewaves, while for open boundary conditions (OBC), {\it all} eigenstates  are localized.
    \item System stability is extremely sensitive to boundary conditions: the BKC is dynamically unstable for PBC, but completely stable for OBC.  
\end{enumerate}
Note that these behaviours are not identical.  While the NHSE implies OBC and PBC will have very different spectra, feature (2) requires something much stronger (namely that the OBC dynamical matrix has purely real spectra).  
Other novel features associated with the BKC include chiral transport and amplification \cite{McDonald_PRX_2018_BKC}, and entanglement phase transitions reminiscent of monitored quantum systems \cite{Lee_PRXQ_2024_BKC_EPT}.  The BKC model is also potentially interesting for applications, both as a phase-sensitive amplifier \cite{Metelmann_arXiv_2025_qlimited_amplification, Villiers_PRXQ_2024_bogoliubov_ampl}, and also as a parametric quantum sensor \cite{McDonald_Natcomm_2020_sensing, Blanchard_arXiv_2025_BKC_sensing_expt}.



What makes the BKC model special?  Can its behaviour be tied to specific symmetries, and if so, can we use these to construct a wider class of models with similar physics?  These are the key questions that motivate the present work. 
We would also like to understand why certain kinds of disorder rapidly degrade the BKC (e.g.~on-site potential terms cause the BKC to lose many of its exceptional features \cite{McDonald_PRX_2018_BKC,Ughrelidze_PRA_2024_interplay_stability, Yokomizo_PRB_2021_nonbloch_bosonic}).   

\begin{figure}[t!]
\centering{}\includegraphics[width=\columnwidth]{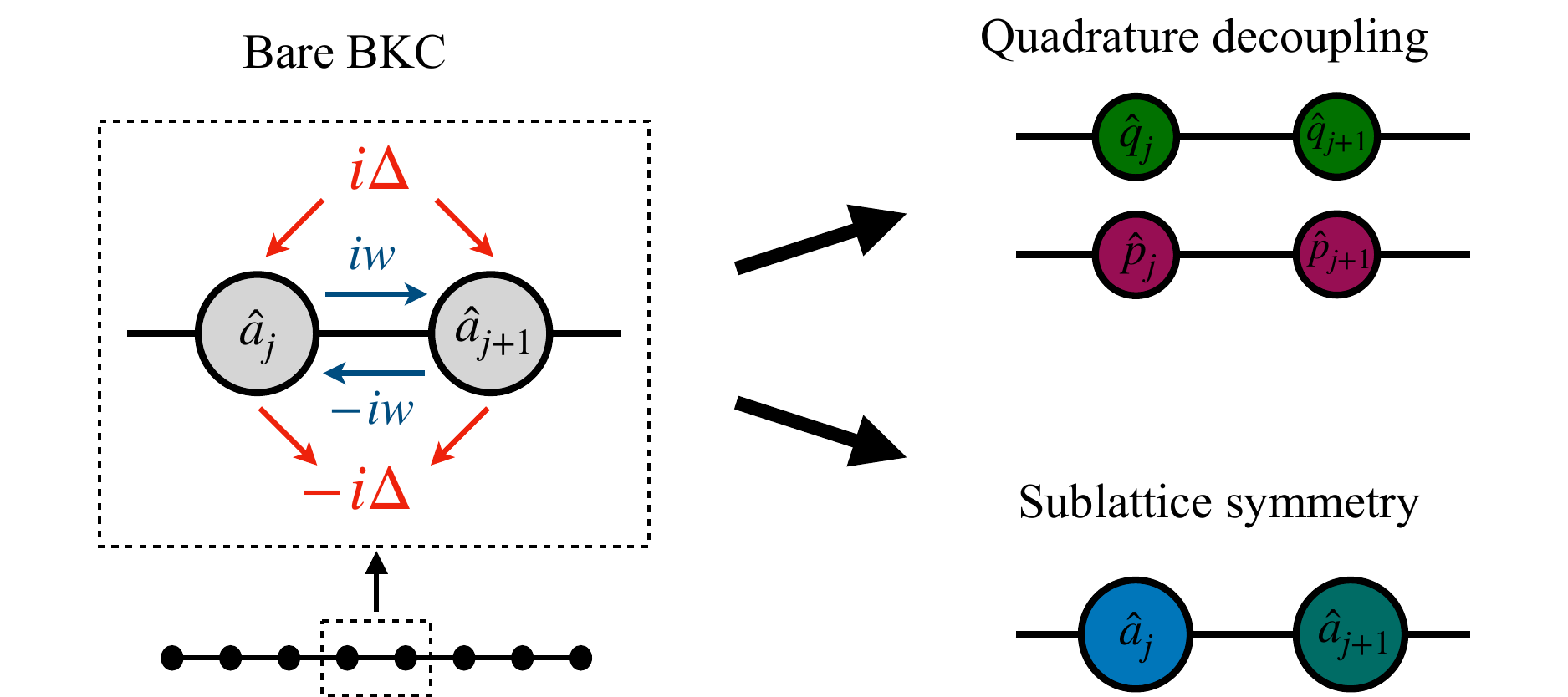}\caption{Left: Schematic of the bosonic Kitaev chain (BKC), a one-dimensional model with hopping $i w$ and pairing $i \Delta$ on each bond.  Despite being Hermitian, it exhibits a number of striking features typically associated with non-Hermitian topology. Right: Two symmetries of the BKC relevant to its exceptional behaviour:  decoupling of $q$ and $p$ quadrature dynamics (which can be mapped to a particle hole type symmetry), and a sublattice or chiral symmetry (which becomes an effective time-reversal symmetry).  More general BKC-like models can be constructed by retaining one or both of these symmetries.   
\label{fig:possible_schematic}}
\end{figure}

In this work, we show that the BKC possesses two distinct symmetries, each of which can be used independently to design new Hermitian bosonic pairing models with non-trivial non-Hermitian topology.  One symmetry is related to a well-known feature of the basic BKC model: when the dynamics is expressed in terms of canonical quadratures, the dynamics of canonical $X$ quadratures decouples from those of $P$ quadratures, giving two copies of a Hatano-Nelson model.  We show that formally, this quadrature-decoupling is the result of an effective particle-hole symmetry, one that can be engineered in more complex models, and even can apply in settings with interactions \cite{lv_arXiv_2025_interacting_BKC}.

Perhaps more interesting is a second symmetry in the basic BKC model that leads to non-trivial consequences, even without any decoupling of quadratures. The relevant symmetry is nothing but simple sublattice (or chiral) symmetry, which is familiar from the classification of standard Hermitian band topology (e.g.~it underlies the celebrated SSH model). Specifically, a quadratic bosonic Hamiltonian $\hat{H} = \hat{H}(\{\hat{a}_j\})$ in 1D is said to have sublattice symmetry if
\begin{equation}
    \hat{H}(\{\hat{a}_j\}) = - \hat{H}(\{(-1)^j \hat{a}_j\}). \label{eq: sublatticesymmetry}
\end{equation}
We show that, somewhat surprisingly, sublattice symmetry in a bosonic pairing Hamiltonian implies that the system's dynamical matrix necessarily possesses an effective time-reversal symmetry (TRS).  This effective TRS has no connection to a physical TRS operation (e.g.~a local anti-unitary symmetry of the original second-quantized $\hat H$).  It however does yield non-trivial non-Hermitian topology \cite{Okuma_PRL_2020_topological_origin}.  A key result of our work is to show how chiral symmetry can be used to construct BKC-like lattice models that have non-trivial behaviour, even without any simple decoupling of canonical quadratures.  We also show that sublattice symmetry can be directly tied to the remarkable dynamical stability properties of BKC-like models (e.g.~stability is intimately tied to boundary conditions).  We also point out that for a slightly generalized version of the BKC model, the effective TRS generated by sublattice symmetry can be connected to a physical time-reversal symmetry of a fermionic non-Hermitian model (namely the well-studied symplectic Hatano-Nelson model \cite{Kawabata_PRX_2019_symmetry_topology_NH, Kawabata_PRX_2023_EPT_from_NHSE}).

The rest of this work is structured as follows. Sec.~\ref{sec:prelims} establishes notation, while Sec.~\ref{sec:symmetry_blitz} reviews the relevant symmetries of the BKC.  We show crucially that the NHSE in the BKC simultaneously realizes two different kinds of skin effects (each typically associated with a distinct kind of model).  For the BKC, these amount to a mechanism related to the decoupling of dynamics in a quadrature basis, versus a mechanism directly tied to sublattice symmetry.  We also show that a variant of the BKC effectively simulates the well-known symplectic Hatano-Nelson (SHN) model. 
In Sec.~\ref{sec:symmetry_and_stability}, we show that beyond the NHSE, sublattice symmetry also protects bosonic systems against instabilities under open boundary conditions. 
In Sec.~\ref{sec:classification}, we use our understanding to classify all translationally invariant 1D quadratic bosonic systems.  We uncover several new BKC-like models which host some combination of features (1) and (2), as well as exotic effects such as phase-sensitive criticality or frequency-selected amplification. 
We include various technical appendices, including 
App.~\ref{sec:interactions}, which discusses how these symmetries can guide the construction of interacting (non-linear) models that retain aspects of the BKC model.  

\begin{table*}[]
\centering
\begin{tabular}{|c|c|c|c|c|c|}
\hline
 \textbf{Hamiltonian terms} & \textbf{Quadratures} & \textbf{Sublattice Symm.} & \textbf{Skin Effect?} & \textbf{Stability?} & \textbf{Examples}  \\ \hline
 $\Delta_d, w_d$, odd $d$ & {\color{OliveGreen} Decoupled}  & {\color{OliveGreen} Unbroken} & {\color{OliveGreen} Single-band + SPSE } &  {\color{OliveGreen} Stable} & Eq.~\eqref{eq:H.BKC.OG} \\ \hline
  $\Delta_d, w_d$, any $d$ & {\color{OliveGreen} Decoupled} & {\color{BrickRed} Broken} & {\color{OliveGreen} SPSE } & {\color{OliveGreen} Stable} & Eq.~\eqref{eq:H_Delta2} \\ \hline
 $\Delta_d, w_d, g_d, \eta_d$, odd $d$ & {\color{BrickRed} Coupled} & {\color{OliveGreen} Unbroken} & {\color{OliveGreen} Single-band }& {\color{BrickRed} Unstable} & Eqs.~\eqref{eq:H.BKC.g1},~\eqref{eq:g3_BKC}  \\ \hline
 $\Delta_d, w_d, g_d, \eta_d$, any $d$ & {\color{BrickRed} Coupled} & {\color{BrickRed} Broken} & {\color{BrickRed} None } & {\color{BrickRed} Unstable} & Eq.~\eqref{eq:H.chem} \\ \hline
\end{tabular}
\caption{Summary of symmetry framework used in this work to understand the NHSE and stability in the BKC and BKC-like models. The terms in the first column correspond to the parameterization in Eq.~\eqref{eq:gen_TI_QBH}.}
\label{table:symmetric_schematic}
\end{table*}

\section{Preliminaries}\label{sec:prelims}

\subsection{Notation}

We review here some preliminary details in order to establish notation and terminology, as well as some of the basic distinguishing features of quadratic Hamiltonians. Importantly, we emphasize the difference between the spectrum of a fermionic system, coming from the BdG Hamiltonian, and the spectrum of a bosonic system, which comes instead from a dynamical matrix, and review the physical interpretation of complex eigenvalues in Hermitian bosonic systems.

For bosonic operators, we will use real-space annihilation operators $\hat{a}_n$ ($n$ being the site index), characterized by commutation relations $[\hat{a}_j, \hat{a}_k^\dagger] = \delta_{j, k}$. Since we will work in 1D, there is no ambiguity with the indexing; for a chain of length $L$ we will number the sites $1, ..., L$. When also imposing periodic boundary conditions, we will identify the site $L+1$ with $1$. When referring to quadrature operators, we will use $\hat{q}_j = (\hat{a}_j + \hat{a}_j^{\dag})/\sqrt{2}$,  $\hat{p}_j = -i(\hat{a}_j -\hat{a}_j^{\dag})/\sqrt{2}$, which are characterized by commutation relations $[\hat{q}_j, \hat{p}_k] = i \delta_{j, k}$. For fermionic operators, we will use $\hat{c}_j$, characterized by anticommutation relations $\{ \hat{c}_j, \hat{c}_k^\dagger  \} = \delta_{j, k}$. For column vectors comprising the respective kinds of operators $\hat{\alpha}_j \in \{\hat{q}_j, \hat{p}_j, \hat{a}_j, \hat{c}_j \}$, we will use bold-face font, i.e. $\hat{\pmb{\alpha}} := (\hat{\alpha}_1, ..., \hat{\alpha}_L)^T$, and $\hat{\pmb{\alpha}}^{\dag} := (\hat{\alpha}_1^{\dag}, ..., \hat{\alpha}_L^{\dag})^T$. When the hat is removed from an operator $\alpha := \langle \hat{\alpha} \rangle$, this refers to a classical expectation value, taken with respect to some state which should be clear from context.

For the second quantized Hamiltonians, we will write a hat, $\hat{H}$, and for associated BdG Hamiltonian we will remove the hat, i.e.
\begin{equation}\label{eq:BdG_H}
    \begin{aligned}
        \hat{H} = \begin{pmatrix}
            (\hat{\mathbf{a}}^{\dag})^T & \hat{\mathbf{a}}^T
        \end{pmatrix} H \begin{pmatrix}
            \hat{\mathbf{a}} \\ \hat{\mathbf{a}}^{\dag}
        \end{pmatrix},
    \end{aligned}
\end{equation}
for bosons, with $\hat{\mathbf{a}} \rightarrow \hat{\mathbf{c}}$ for the fermionic case. For bosons, we may sometimes favour writing in the quadrature basis. To convert between the two, we use the unitary matrix, 
\begin{equation}
    \begin{pmatrix}
        \hat{a}_j \\ \hat{a}_j^{\dag}
    \end{pmatrix} = \frac{1}{\sqrt{2}} \begin{pmatrix}
        1 & i \\
        1 & -i
    \end{pmatrix} \begin{pmatrix}
        \hat{q}_j \\ \hat{p}_j
    \end{pmatrix} =: u_1 \begin{pmatrix}
        \hat{q}_j \\ \hat{p}_j
    \end{pmatrix}.
\end{equation}
Writing $U = \oplus_{n=1}^L u_1$, we then denote the Hamiltonian in the quadrature basis with superscripts, $H^{\rm qp}$, so 
\begin{equation}
    H^{\rm qp} := U^{\dag} H U.
\end{equation}
We reserve subscripts for denoting different models we may look at. 

When discussing PBC, we will be dealing exclusively with translationally invariant Hamiltonians, for which we use the Fourier transformation, 
\begin{equation}
    \hat{a}_j = \frac{1}{\sqrt{N}} \sum_{k} e^{ i k j} \hat{a}[k],
\end{equation}
where we write $\sum_{k}$ in this and future equations to mean the sum over $k$ goes over $k = \frac{2 \pi }{N}, \frac{4 \pi }{N} ..., \frac{2 N \pi}{N}$, and we identify operators in the Fourier basis with a square bracket argument. For the BdG Hamiltonians in the Fourier basis, we will use $H(k)$, eg. 
\begin{equation}
    \hat{H} = \sum_{k} \begin{pmatrix}
        \hat{a}[k]^{\dag} & \hat{a}[-k]
    \end{pmatrix} H(k) \begin{pmatrix}
        \hat{a}[k] \\ \hat{a}[-k]^{\dag}
    \end{pmatrix},
\end{equation}
where the Hamiltonian is block-diagonalized due to translational invariance. 

Finally, for quadratic fermionic systems, the dynamics are determined directly by $H$, and we refer to the eigenvalues of $H$ as its spectrum. In contrast, for quadratic bosonic systems, the dynamics are given instead by the \textit{dynamical matrix}. For dynamical matrices of bosonic systems, we will default to the quadrature basis; the dynamical matrix $M$ is then the generator of the equations of motion, 
\begin{equation}\label{eq:bosonic_eom}
    \begin{aligned}
        \frac{d}{dt} \begin{pmatrix}
            \hat{\mathbf{q}} \\ \hat{\mathbf{p}}
        \end{pmatrix} =: -i M^{\rm qp} \begin{pmatrix}
            \hat{\mathbf{q}} \\ \hat{\mathbf{p}}
        \end{pmatrix}.
    \end{aligned}
\end{equation}
For quadratic (Gaussian) bosonic systems, the dynamical matrix is well-defined since the equations of motion are linear. The dynamical matrix is readily obtained from the BdG Hamiltonian via
\begin{equation}
    M = i Z H, \qquad Z := \bigoplus_{n=1}^N \sigma_z,
\end{equation}
or in the quadrature basis as 
\begin{equation}
    M^{\rm qp} = i \Omega H^{\rm qp}, \qquad \Omega := \bigoplus_{n=1}^N \begin{pmatrix}
        0 & 1 \\ -1 & 0
    \end{pmatrix},
\end{equation}
where $Z, \Omega$ are to be understood as the symplectic form encoding the commutation relations of the bosonic operators (in either basis). The two are related by $Z = U^{\dag} \Omega U$. For more pedagogical details, see Ref.~\cite{Serafini_2023_QCV_textbook}. 

As a consequence, for bosonic systems, we will refer to the eigenvalues of the dynamical matrix $M$ as its spectrum. In contrast to fermionic systems, the spectrum of a quadratic bosonic system is in general complex, even when the Hamiltonian is completely Hermitian. The complexity of the spectrum encodes the infinite dimensional nature of Fock space; eigenmodes whose eigenvalues have imaginary part are said to be dynamically unstable, as their occupation is amplified unboundedly, as can be observed in Eq.~\eqref{eq:bosonic_eom}. Such instabilities can only occur in the presence of pairing/squeezing terms $\hat{a}_i \hat{a}_j, \hat{a}_i^{\dag} \hat{a}_j^{\dag}$. In experiment, this amplification is ultimately cut off by nonlinearities, so as to remain physical. For stable systems (with or without squeezing), one may understand the topology of quadratic bosonic systems similarly to Hermitian free fermionic systems \cite{Chaudhary_PRB_2021_bosonic_band_topology}. In the presence of instability, one must bring in topology of non-Hermitian free fermionic systems \cite{Kawabata_PRX_2019_symmetry_topology_NH, Gong_PRX_2018_symmetry_topology_NH_old}.

\begin{table}[]
\begin{tabular}{|r|l|}
\hline
 BKC & Bosonic Kitaev chain \\ \hline
  NHSE & Non-Hermitian skin effect  \\ \hline
  SLS& Sublattice symmetry\\ \hline
  PHS & Particle-hole symmetry \\ \hline
  aPHS & Automatic particle-hole symmetry \\ \hline
  qPHS & Quadrature particle-hole symmetry \\ \hline
  TRS & Time-reversal symmetry \\ \hline
  SPSE & Symmetry-protected skin effect \\ \hline
  OBC & Open boundary conditions \\ \hline
  PBC & Periodic boundary conditions \\ \hline
\end{tabular}
\caption{Table of acronyms used in this work}
\label{table:acronyms}
\end{table}

\subsection{Topology and the NHSE}\label{ssec:tops_nhse}

In this section, we briefly recap the essential ingredients of the connection between band topology and NHSE. For details, we refer readers to Ref.~\cite{Okuma_PRL_2020_topological_origin}, or the more general reviews Ref.~\cite{Okuma_2022_NH_review, Ashida_2020_NH_review, Bergholtz_2021_RMP_NH_review, Lin_2023_frontiers_topological_NH_review}. 

Recall that a single-band model characterized by a matrix $H$ under PBC (which may be e.g.~the Hamiltonian of a free fermionic model, or the dynamical matrix of any physical problem) is said to have a \textit{point gap} with respect to a reference point $E \in \mathbb{C}$ if the spectrum of $H$ does not cross $E$ \cite{Kawabata_PRX_2019_symmetry_topology_NH}. One may accordingly define a winding number,
\begin{equation}\label{eq:single-band_winding}
    W(E) = \int_0^{2\pi} \frac{dk}{2 \pi i} \frac{d}{dk} {\rm log} \, {\rm det} (H(k) - E),
\end{equation}
which diagnoses the point gap, i.e.~$H$ has a point gap with respect to $E$ if and only if $W(E) \neq 0$. For such models, 
having a non-zero winding number is in one-to-one correspondence with the existence of a NHSE. Concretely, consider an eigenstate $|\Psi_E \rangle$ of the system under OBC. Suppose that using the PBC spectrum, one evaluates $W(E) \neq 0$.  Then the results of Ref.~\cite{Okuma_PRL_2020_topological_origin, Zhang_PRL_2020_GBZ_BZ_origin_NHSE} tell us that $| \Psi_E \rangle$ is a skin mode -- that is, a mode that is exponentially localized to a boundary of the system. In this work, we will refer to this kind of a skin effect as a \textit{single-band skin effect}. As an example, the canonical Hatano-Nelson model hosts a single-band NHSE \cite{Yao_PRL_2018_NHSE_OG_2, Martinez_PRB_2018_NHSE_OG_3, Hatano_1997_vortexpinning, Hatano_1996_HNmodel}.

NHSE in single-band non-Hermitian models are the easiest to understand; unfortunately, due to the doubling of the degrees of freedom in a bosonic BdG Hamiltonian (see Eq.~\eqref{eq:BdG_H}), the dynamical matrix $M$ of a quadratic bosonic Hamiltonian is {\it never} a single-band model (see Sec.~\ref{sec:symmetry_blitz} for more details). Furthermore, evaluating Eq.~\eqref{eq:single-band_winding} for any bosonic dynamical matrix (set $H(k) \rightarrow M(k)$) yields $W(E) = 0$ for any value of $E$.  Naively one might conclude that skin effects in Hermitian bosonic systems are impossible.  This is known to be incorrect (see e.g.~\cite{Kawabata_PRX_2019_symmetry_topology_NH, Busnaina_natcomm_2024_BKC_expt, Slim_nature_2024_BKC_expt}), suggesting that we must understand mechanisms for the NHSE that go beyond the simple single-band mechanism described above.  As we now show, there are two distinct new mechanisms that are relevant and that can generate a NHSE.   

First, in some cases the linear dynamics of the system decouples in a particular basis of quadratures (i.e.~the dynamics of $\hat{q}$ quadratures is completely independent of those of $\hat{p}$ quadratures and vice versa, for some choice of canonically conjugate $q$ and $p$).  We show below that this can be directly tied to a particular particle-hole symmetry of the dynamical matrix $M(k)$.  This quadrature decoupling allows us to view our system as two completely decoupled single-band systems.  We can thus define a winding number as defined in Eq.~\eqref{eq:single-band_winding} for each of the two decoupled quadratures \footnote{One may object that such decoupling is always possible given an appropriate change of basis. The important thing here is that the definition of the quadratures is the same on each site, i.e. we have a local change of basis that is $k$-independent.}.
We may understand the phenomenology of such models by simply regarding them as separate instances of a single-band model  in the spirit of the single-band skin effect.

Alternately, a richer form of the NHSE may arise from the presence of more non-trivial symmetries. This is a \textit{symmetry-protected skin effect} (SPSE). In particular, suppose a dynamical matrix $M$ hosts a time-reversal symmetry (TRS) of the transpose form,
\begin{equation}\label{eq:TRS_def}
    T M^T(k) T^{-1} = M(-k),
\end{equation}
where $T$ is a unitary matrix satisfying $T T^* = -1$, which we refer to as the symmetry operator. This symmetry furnishes a $\mathbb{Z}_2$ invariant $\nu(E) \in \{0, 1\}$, again with respect to a reference point $E \in \mathbb{C}$, 
\begin{equation}\label{eq:TRS_winding}
\begin{aligned}
    &(-1)^{\nu(E)} := \\
    &{\rm sgn} \left[ P[M] \exp \left( - \frac{1}{2} \int_{k=0}^{k=\pi} d \log {\rm det} \, \left( (M(k) - E) T\right) \right)  \right],
\end{aligned}
\end{equation}
where 
\begin{equation}
    P[M] = \frac{{\rm Pf}[(M(\pi) - E) T]}{{\rm Pf}[(M(0) - E) T]}.
\end{equation}
A non-zero value of this invariant also signals a NHSE, giving a generalization of the single band NHSE: eigenstates of $M$ under OBC at energy $E$ are localized when $\nu(E) = 1$. When clear from context, we will also refer to Eq.~\eqref{eq:TRS_winding} as a winding number, and when $\nu(E) = 1$, we say that we have an SPSE arising from TRS. 

Finally, we note that there is another SPSE proposed in Ref.~\cite{Okuma_PRL_2020_topological_origin}, arising directly from particle-hole symmetry (PHS). Unfortunately, the relevant SPSE is weaker in this case, as it only yields a winding number with respect to $E = 0$, analogous to the Hermitian case \cite{Bernevig_Hughes_2013}. While the models we examine later may exhibit particle-hole symmetries, the SPSE arising from PHS will not be directly useful to us. Compared to a PHS, in the non-Hermitian setting, a TRS of the above form is much more powerful, as defines invariants for arbitrary $E$'s.

\section{Symmetries of the bosonic Kitaev chain}\label{sec:symmetry_blitz}

\subsection{Generic quadratic bosonic Hamiltonians and built-in symmetries}

We start with a general translationally-invariant quadratic Hamiltonian describing bosons on a 1D lattice: 
\begin{equation}\label{eq:gen_TI_QBH}
    \begin{aligned}
        \hat{H} = \frac{1}{2} \sum_{d \geq 0} \sum_{j} &\left( (g_d + i w_d) \hat{a}^{\dag}_{j+d} \hat{a}_j \right. \\
        &\left. + (i \Delta_d + \eta_d) \hat{a}^{\dag}_{j+d} \hat{a}_j^{\dag} + {\rm H.c.} \right),
    \end{aligned}
\end{equation}
where $d \geq 0$ indexes a coupling range,
and the parameters $g_d, w_d, \Delta_d, \eta_d$ are real. Hermiticity forces $w_0 = 0$.
Recall that the standard BKC model (Eq.~\eqref{eq:H.BKC.OG}) is specified by keeping only $w_1, \Delta_1$ nonzero.

In Fourier space, this takes the form, 
\begin{equation}
    H(k) = 
    \begin{pmatrix}
        A(k) & B(k) \\ B(-k)^* & A(-k)
    \end{pmatrix},
\end{equation}
where 
\begin{equation}\label{eq:AB_parameterization}
    \begin{aligned}
        A(k) &= \sum_d \left( w_d \sin(k d ) - g_d \cos(kd)  \right), \\
        B(k) &= \sum_d \left(  i \Delta_d - \eta_d \right) \cos(kd).
    \end{aligned}
\end{equation}
The dynamical matrix in momentum space is then
\begin{equation}
    M(k) = \sigma_z H(k) = \begin{pmatrix}
        A(k) & B(k) \\ -B(-k)^* & -A(-k)
    \end{pmatrix},
\end{equation}
where we note that $B(k) = B(-k)$.

For any choice of parameters, our model always has several built-in symmetries, which we refer to as `automatic' symmetries. The first is a particle-hole symmetry (PHS) due to the doubling of the degrees of freedom when going to the BdG space, 
\begin{equation}\label{eq:ogPHS}
    \sigma_x M(-k)^* \sigma_x = - M(k). 
\end{equation}
The second automatic symmetry is pseudo-Hermiticity, where we have 
\begin{equation}
    \sigma_z M(k)^{\dag} \sigma_z = M(k).
\end{equation}
This stems from the Hermiticity of Eq.~\eqref{eq:gen_TI_QBH}. Finally, we may combine these two to obtain yet another particle-hole symmetry, this time of the form, 
\begin{equation}\label{eq:aPHS}
    \sigma_y M(-k)^T \sigma_y = - M(k).
\end{equation}
For easy reference, and to contrast with other particle-hole symmetries that will turn up later, we refer to the symmetry denoted by Eq.~\eqref{eq:aPHS} as aPHS (automatic). We note it is more standard to write PHS in terms of complex conjugation, like in Eq.~\eqref{eq:ogPHS}. The version of PHS with a transpose, like in Eq.~\eqref{eq:aPHS} is equivalent to the former for Hermitian systems, but is an independent symmetry in the non-Hermitian case. 

Remarkably, in the non-Hermitian context, it is only the PHS of the latter form that defines a winding number for the NHSE. One could directly attempt to use these automatic symmetries to define a meaningful topological winding number. However, as shown in Refs.~\cite{Busnaina_natcomm_2024_BKC_expt, Kawabata_PRX_2019_symmetry_topology_NH, Wan_PRL_2023_squeezing_induced_NHSE}, in the absence of additional symmetries, the resulting winding numbers  will always be trivial with respect to a point gap.

\subsection{The basic bosonic Kitaev chain model}

We now specialize to the standard BKC model, and quickly review some of its key symmetries.  While many of these properties have been discussed in isolation in previous works (see e.g.~\cite{Busnaina_natcomm_2024_BKC_expt, Kawabata_PRX_2019_symmetry_topology_NH, Fortin_PRB_2025_topological_BKC_loss, Wan_PRL_2023_squeezing_induced_NHSE, Flynn_NJP_2020_deconstructing_NH_dynamics}), our discussion here serves to unify these ideas and bring them together in one place. It also will explicitly separate two easily-conflated but ultimately distinct sources of the skin effect in the BKC.

The Hamiltonian of the basic BKC is given by Eq.~\eqref{eq:H.BKC.OG}.
This corresponds to the more general Hamiltonian in Eq.~\eqref{eq:gen_TI_QBH} with the parameter choices $w_d = w \delta_{d, 1}$, $ \Delta_d = \Delta \delta_{d, 1}, g_d = \eta_d = 0$.  The momentum-space dynamical matrix of the basic BKC model takes the form:
\begin{equation}
    M_{\rm BKC}(k) = w \sin k + i \Delta \cos k \sigma_x
\end{equation}

The dynamical matrix $M$ has two bands, and as discussed in Sec.~\ref{ssec:tops_nhse}, we thus have two routes for diagnosing an NHSE via a non-zero winding number: either we reduce the model to two decoupled single-band models (yielding a single-band skin effect), or we apply the notion of a symmetry-protected skin effect (SPSE). As we show both these invariants are non-zero for the basic BKC model, and predict a NHSE.  This coinciding of different NHSE mechanisms is unique to the BKC, and will not be true in more general models. The fact that they coincide here is a consequence of the abundance of symmetry in the basic BKC model.

\subsection{Decoupling of quadratures as an additional particle-hole symmetry} 

It is well-known that the dynamics of the BKC reduces to two decoupled Hatano-Nelson models  \cite{McDonald_PRX_2018_BKC}. This is readily seen by going to the quadrature basis, which diagonalizes the dynamical matrix:
\begin{equation}
\begin{aligned}
    &M^{\rm qp}_{\rm BKC}(k) 
    = u_1^{\dag} M_{\rm BKC}(k) u_1 \\
    &= \begin{pmatrix}
        w \sin(k)  + i \Delta \cos(k) & 0 \\ 0 & w \sin(k) -  i \Delta \cos(k)
    \end{pmatrix}.
\end{aligned}
\end{equation}
Each diagonal element gives the dispersion of a single Hatano-Nelson chain, with opposite signs in $\Delta$.

In terms of topological classification, the above decoupling can be formalized in terms of a second particle-hole symmetry \cite{Wan_PRL_2023_squeezing_induced_NHSE, Fortin_PRB_2025_topological_BKC_loss},
\begin{equation}
\label{eq:M_BKC}
\begin{aligned}
    M_{\rm BKC}(-k)^* 
    = - w \sin k - i \Delta \cos k \sigma_x 
    = - M_{\rm BKC}(k),
\end{aligned}
\end{equation}
where the symmetry operator is the identity. We note that this particle-hole symmetry uses complex conjugation, whereas the aPHS uses a transpose -- recall that these operations crucially differ in the non-Hermitian setting. In the position basis, the corresponding transformation is simply complex conjugation, i.e.~
\begin{align}
M_{\rm BKC}^*=-M_{\rm BKC},
\end{align}
which forces the parameters of the Hamiltonian to be purely imaginary (something that is of course true for the standard BKC model).
We refer to this as qPHS (quadrature). We now show that the presence of qPHS along with the built-in symmetries of our dynamical matrix guarantees a decoupling of the dynamics into two independent single band models. 


Consider a generic dynamical matrix that satisfies qPHS.  Starting with this condition and using the two built-in symmetries of any bosonic model, it follows that $M(k)$ must commute with $\sigma_x$:
\begin{equation}\label{eq:qPHS_equiv}
    \begin{aligned}
        M(k) &= - M(-k)^* && \mbox{ (qPHS) } \\
        &= - \sigma_z M(-k)^T \sigma_z && \mbox{ (pseudo-Hermiticity) } \\ 
        &= \sigma_y \sigma_z M(k) \sigma_z \sigma_y && \mbox{ (aPHS) } \\
        &= \sigma_x M(k) \sigma_x.
    \end{aligned}
\end{equation}
It follows that $M(k)$ must be diagonal in the basis of $\sigma_x$, which in turn means that the dynamics of the standard $q,p$ quadratures decouple.  The above set of equalities also implies that if the dynamics is diagonal in the quadrature basis, then the model must have qPHS, so the two notions are equivalent for quadratic bosonic Hamiltonians. Note that more generally, one may have a PHS of the form $M(k) = - U M(-k)^* U^{-1}$ for which this discussion still holds up to a local change of basis. Here, $U$ is a unitary matrix independent of $k$. On the second-quantized level, this corresponds to a uniform phase $\hat{a}_j \rightarrow e^{i\theta} \hat{a}_j$.  Equivalently, one might say we have two internal symmetries of the same type (PHS$^\dagger$ \cite{Kawabata_PRX_2019_symmetry_topology_NH}), which combine to give a unitary symmetry; the symmetry analysis can then be done within the sub-blocks, which amounts to a reduction to single-band models. We discuss this more general perspective in Sec.~\ref{ssec:BKC-38}.

With this reduction to two single-band models, for any point $E \in \mathbb{C}$ on the complex plane, we may now define a \textit{pair} of winding numbers. We use Eq.~\eqref{eq:qPHS_equiv} to write 
\begin{equation}
    M(k) = \begin{pmatrix}
        M_q(k) & 0 \\ 0 & M_p(k)
    \end{pmatrix},
\end{equation}
where we further note that pseudo-Hermiticity implies that $M_q(k)^* = M_p(k)$. Treating $M_q, M_p$ as separate single-band models, we then define
\begin{equation}
\label{eq:doubled_winding}
    W_{q/p}(E) = \frac{1}{2\pi i} \int_0^{2 \pi} dk \frac{\partial}{\partial k} {\rm log} \mbox{ det} (M_{q/p}(k) - E),
\end{equation}
where we note that in general, $W_q(E) \neq W_p(E)$. We emphasize that the validity of this number in diagnosing the skin effect comes from the fact that $M_q, M_p$ individually host skin effects, rather than being protected by qPHS. However, the qPHS does provide a relation between the two winding numbers,
\begin{equation}
    W_q(E) = W_p(E^*).
\end{equation}
Furthermore, we note that for the bare BKC, the two winding numbers are always equal, as the two bands coincide. This is not true in general, as we will see in subsequent sections.

\subsection{Effective TRS from sublattice symmetry and symmetry-protected skin effects}

The skin effect enabled by quadrature decoupling is not very different from the single-band physics of the original Hatano-Nelson model. We will now do as promised in the introduction, and show that a different, richer skin effect is also present in the BKC. 

At first glance, the momentum-space dynamical matrix of the basic BKC model (c.f.~Eq.~\eqref{eq:M_BKC}) does not possess a symmetry of the form Eq.~\eqref{eq:TRS_def}, i.e.~an additional symmetry that would yield a SPSE. To obtain such a symmetry, we take a rather surprising route. First, we observe that the bare BKC fulfills a sublattice symmetry (SLS). On the level of second quantized operators, this is the symmetry given by 
\begin{equation}\label{eq:sublattice_req}
    \begin{aligned}
        \hat{a}_n &\rightarrow \hat{a}_n (-1)^n \\
        \hat{H}_{\rm BKC} &\rightarrow - \hat{H}_{\rm BKC}.
    \end{aligned}
\end{equation}

Recall that one standard way to motivate a chiral symmetry in Hermitian systems is as a combination of time reversal and particle-hole symmetry \cite{AZ_classes}.  Similarly, a TRS can arise from a combination of PHS and chiral symmetry (which can be realized through a SLS). The same thing happens in non-Hermitian systems \cite{Kawabata_PRX_2019_symmetry_topology_NH}, although there is extra subtlety as there are two varities of TRS (defined either via transpose or conjugatation operations). It will be instructive for us to step through how this TRS is obtained, which will allow us to understand the SPSE in the BKC. 

In this work, we focus on models with full translational symmetry (see Eq.~\eqref{eq:gen_TI_QBH}). Despite this, it will be useful to treat the system as effectively having a two-site unit cell. To wit, we pass to the folded zone picture, where we combine the Fourier components $\hat{a}[k], \hat{a}^{\dag}[-k]$ with their $\pi + k$ counterparts as a 4-component column vector, as these are the components coupled by the SLS. The $4 \times 4$ dynamical matrix $\mathcal{M}[k]$ in our folded-zone representation then has form 
\begin{equation}
\begin{aligned}
    \mathcal{M}[k] = 
&\begin{pmatrix}
        M(k) & 0 \\ 0 & M(k + \pi)
    \end{pmatrix}, 
\end{aligned}
\end{equation}
where each entry of the above is a two-by-two block.

At first glance, it may seem strange to pretend the model has less translational symmetry than it actually does. In one case, we know this perspective has been fruitful -- for the bare BKC, there exists a gauge transformation that makes the Hamiltonian real at the expense of doubling the unit cell \cite{McDonald_PRX_2018_BKC}. Similarly, for us the folded-zone representation will reveal a symmetry relevant to the skin effect. In this picture, there is a simple matrix representation of the SLS via a matrix $S$ that acts as a Pauli $\sigma_x$ matrix in this space, i.e.~$S \simeq \sigma_x \otimes I$.  Explicitly, we have: 
\begin{equation}
    S \begin{pmatrix}
        \hat{a}[k] \\ \hat{a}^{\dag}[-k] \\ \hat{a}[k + \pi] \\ \hat{a}^{\dag}[\pi-k]
    \end{pmatrix} =  \begin{pmatrix}
        \hat{a}[k + \pi] \\ \hat{a}^{\dag}[\pi-k] \\ \hat{a}[k] \\ \hat{a}^{\dag}[-k]
    \end{pmatrix}.
\end{equation}
We see then that $S$ satisfies the needed conditions for a chiral symmetry, namely:  
\begin{equation}
    S 
    \mathcal{M}[k] S^{-1} = (\sigma_x \otimes I) \mathcal M[k]  (\sigma_x \otimes I) = 
    -\mathcal{M}[k]
    \label{eq:FoldeZoneChiral}
\end{equation} 

For a generic dynamical matrix with SLS, we can then combine Eq.~(\ref{eq:FoldeZoneChiral}) with Eq.~\eqref{eq:aPHS} (i.e.~form the product of  SLS with aPHS), to obtain a formal transpose-type TRS symmetry of our system, i.e. a mapping between the folded-zone dynamical matrix $(\mathcal{M}[k])^T$ and $\mathcal{M}[-k]$:
\begin{equation}\label{eq:kTRS_eq}
\begin{aligned}
    (\sigma_x \otimes \sigma_y) 
    \mathcal{M}^T[k]
    (\sigma_x \otimes \sigma_y) 
     = 
    \mathcal{M}[-k]. 
\end{aligned}
\end{equation}
Note further that $(\sigma_x \otimes  \sigma_y) (\sigma_x \otimes \sigma_y)^* = - I \otimes I$, so this fulfills Eq.~(\ref{eq:TRS_def}), albeit at the cost of doubling the unit cell. 
The upshot is that we have shown that for our quadratic pairing bosonic Hamiltonian, sublattice symmetry automatically implies a formal transpose-type time reversal symmetry.



By evaluating the initial set of conditions with $T = i \sigma_y$, and noting that (via straightforward computation), 
\begin{equation}
\begin{aligned}
    {\rm det} \left( (M(k) - E)  \sigma_y \right) &={\rm det} \left( (M(k) - E) \right), \\
    P[M] &= -1.
\end{aligned}
\end{equation}
we find that this in fact lets us define a winding number $\nu_{\rm sl}$ via, 
\begin{equation}
\label{eq:sublattice_winding}
\begin{aligned}
    &(-1)^{\nu_{\rm sl}(E)} := \\
    &{\rm sgn} \left[ - \exp \left( - \frac{1}{2} \int_{k=0}^{k=\pi} d \log {\rm det} \left( (M(k) - E)\right) \right)  \right],
\end{aligned}
\end{equation}
which characterizes the SPSE.

Going forward, it may also be convenient to make use of this TRS without zone-folding.  We note that the above TRS condition also implies a constraint on our original single-band $2 \times 2$ dynamical matrix that relates momenta $k$ and $\pi - k$:
\begin{equation}\label{eq:BKC_SPSE}
    \sigma_y M(k)  \sigma_y = M(\pi - k),
\end{equation}
with $ \sigma_y \sigma_y^* = -I$. 

\bigskip



\subsection{Effective TRS is not physical TRS}
\label{lsubsec:TRSNotTRS}


The previous section showed that a bosonic pairing lattice model with sublattice symmetry {\it automatically} has an effective time-reversal symmetry at the level of its dynamical matrix, as expressed in Eq.~\eqref{eq:kTRS_eq}.  It is of course tempting to think that there must be a connection to a physical time-reversal symmetry of the original model, that is a local anti-unitary symmetry of the second quantized Hamiltonian (see e.g.~\cite{Koch_PRA_2010_TRS}).  We call this more familiar kind of TRS a physical TRS (pTRS) in what follows.  It is well known that the basic BKC model possesses pTRS \cite{McDonald_PRX_2018_BKC,Wanjura_2023_quadrature_nonreciprocity}, in addition to the effective TRS that follows from sublattice symmetry.  We show below that this is coincidental: these two kinds of TRS are not equivalent, and do not coincide in more general models.   


A necessary and sufficient condition for a bosonic system to satisfy pTRS is the existence of a local gauge transformation
\begin{equation}
a_j \rightarrow a_j e^{i\phi_j},
\end{equation}
such that the Hamiltonian becomes entirely real in the transformed basis~\cite{Koch_PRA_2010_TRS,Wanjura_2023_quadrature_nonreciprocity}.
For the bare BKC, one readily verifies that choosing
\begin{equation}
\phi_j = j\frac{\pi}{2}
\end{equation}
renders the Hamiltonian real \cite{McDonald_PRX_2018_BKC}. The model therefore possesses pTRS (albeit at the expense of doubling the unit cell).

More generally, pTRS and the effective TRS defined in Eq.~\eqref{eq:kTRS_eq} are independent notions. For example, pTRS may be broken while the effective TRS remains intact. This occurs upon adding a next-to-next-nearest-neighbor hopping term to the Hamiltonian
\begin{equation}
    g_3\sum_j a_{j+3}^{\dagger}a_j,
\end{equation}
see also Sec.~\ref{subsec:with_g3}.
Conversely, one may break the effective TRS while preserving pTRS, for instance by introducing a real on-site detuning (or chemical potential) term,
\begin{equation}
    g_0\sum_j a_j^{\dagger}a_j,
\end{equation}
see also Sec.~\ref{subsec:with_g0}.
We thus conclude that these two kinds of TRS should be regarded as distinct.  The effective TRS of Eq.~\eqref{eq:kTRS_eq} is thus best regarded as synonymous with sublattice symmetry in bosonic pairing Hamiltonians.

In this work, we focus primarily on the effective TRS, as it is the symmetry most directly relevant to the physics of BKC-like models, as will become clear in the following sections.

\subsection{Synopsis: BKC skin effects and symmetry}

The upshot of this section's discussion has been to show that the skin effect in the BKC can be understood in two distinct ways, each stemming from truly distinct symmetries. One route to understanding the BKC skin effect is to view it as being enabled (and protected) by qPHS (c.f.~Eq.~\eqref{eq:M_BKC}). However, the topological invariant here does not directly arise from qPHS, but stems in fact from the notion of \textit{quadrature decoupling}, something that maps our system to an effective single-band model. Skin effects stemming from qPHS thus have a more limited nature, as they are equivalent (apart from doubling) to what one could obtain in a single-band non-Hermitian model.  

Alternatively, we demonstrated that the BKC skin effect can also be understood as a being protected by \textit{sublattice symmetry}.  In contrast to skin effects solely stemming from qPHS, this kind of skin effect is a true symmetry-protected skin effect (SPSE).  This kind of skin effect is non-trivially enabled by the presence of symmetry, and
in general cannot be reduced to the physics of a simple single-band model. 

In the bare BKC, there is of course just a single skin effect, one that can be understood either using qPHS or sublattice symmetry.  Recalling the winding numbers defined in Eqs.~\eqref{eq:doubled_winding} and \eqref{eq:sublattice_winding}, for the bare BKC we have $W_q(E) \neq 0 \iff W_p(E) \neq 0 \iff \nu_{\rm sl}(E) \neq 0$ for any $E$. We stress however that the equivalence of these views of the skin effect in the bare BKC will not be true in more general models, as in general, the relevant symmetries need not coincide. Stated explicitly, the two relevant symmetries (sublattice symmetry and quadrature decoupling) can be 
{\it independently} broken or unbroken. We explore this more in Sec.~\ref{sec:classification}.  We note that in general that sublattice symmetry is much more natural symmetry to emerge in a lattice model, whereas qPHS (quadrature decoupling) requires more fine tuning.  

\subsection{Connection to the symplectic Hatano-Nelson model}

We now take a short detour to draw the connection between an oft-studied variant of the BKC and an oft-studied non-Hermitian fermionic model. As mentioned, the TRS of Eq.~\eqref{eq:BKC_SPSE} does not seem to correspond to a physical notion of TRS in the quadratic bosonic model. Nevertheless, we will now show that there is an equivalent fermionic model for which this translates into a bona fide TRS. Let us consider a BKC with additional nearest-neighbor hopping terms with real coefficients. Following Eq.~\eqref{eq:gen_TI_QBH}, we will use the notation $g_d$ to denote a $d$-th nearest-neighbor hopping term, and refer to this model as the ``$g_1$-BKC". This presents our first example of a BKC-like model. The Hamiltonian \cite{McDonald_PRX_2018_BKC} is, 
\begin{equation}
\label{eq:H.BKC.g1}
\begin{aligned}\hat{H}_{g_1} & = \hat{H}_{\rm BKC} + \frac{1}{2}\sum_{j}g_1 \left(\hat{a}_{j+1}^{\dag}\hat{a}_{j}+\text{H.c}.\right),\end{aligned}
\end{equation}
where $\hat{H}_{\rm BKC}$ is as defined in Eq.~(\ref{eq:H.BKC.OG}). The $g_1$ term breaks qPHS and thus this model does not enjoy quadrature decoupling. However, the $g_1$ term does respect sublattice symmetry, so this model retains the SPSE. Notably, the $g_1$-BKC hosts both a skin effect phase when $g_1 < \Delta$ and a trivial phase when $g_1 > \Delta$. In addition to a localization transition, this also hosts a post-selection free entanglement phase transition \cite{Lee_PRXQ_2024_BKC_EPT}. 

The dynamical matrix in momentum space is,
\begin{equation}\label{eq:g1_dynamical}
    M_{g_1}(k) = w \sin k + i \Delta \cos k \sigma_x - g_1 \cos k \sigma_z.
\end{equation}
In quadrature space, we may think of the $g_1$ term as coupling the two effective Hatano-Nelson chains in the BKC. To see this, we write the Heisenberg EOMs of the quadratures,
\begin{equation}\label{eq:g1.BKC.EOM}
\begin{aligned}
\frac{d}{dt}  \hat{q}_{j}   &=
\frac{w + \Delta}{2}\hat{q}_{j-1}-\frac{w - \Delta}{2}\hat{q}_{j+1}+\frac{g_1}{2}(\hat{p}_{j-1}+\hat{p}_{j+1}),\\
\frac{d}{dt}\hat{p}_{j} & =\frac{w - \Delta}{2}\hat{p}_{j-1}-\frac{w + \Delta}{2}\hat{p}_{j+1}-\frac{g_1}{2}(\hat{q}_{j-1}+\hat{q}_{j+1}).
\end{aligned}
\end{equation}
Since the two chains are non-reciprocal in different directions, when $g_1$ is strong enough, it causes the breakdown of the skin effect.

On the other hand, the symplectic Hatano-Nelson (SHN) model \cite{Kawabata_PRX_2019_symmetry_topology_NH, Kawabata_PRX_2023_EPT_from_NHSE} is a fermionic non-Hermitian model that also hosts a localization transition and an entanglement phase transition \cite{Kawabata_PRX_2023_EPT_from_NHSE}. The SHN explicitly describes two Hatano-Nelson chains, coupled by a spin-orbit term. Intentionally reusing notation, the Hamiltonian is,
\begin{equation}
    \begin{aligned}\label{eq:H.SHN}
    &\hat{H}_{\rm SHN} = \\
    & - \frac{1}{2} \sum_{j} \left[ \hat{c}_{\uparrow, j + 1}^{\dag} \frac{w + \Delta}{2} \hat{c}_{\uparrow, j} + \hat{c}_{\uparrow, j}^{\dag} \frac{w - \Delta}{2} \hat{c}_{\uparrow, j+1} \right]  \\
    & - \frac{1}{2} \sum_{j} \left[ \hat{c}_{\downarrow, j + 1}^{\dag} \frac{w - \Delta}{2} \hat{c}_{\downarrow, j} + \hat{c}_{\downarrow, j}^{\dag} \frac{w + \Delta}{2} \hat{c}_{\downarrow, j+1} \right]  \\
    & - \frac{1}{2}\sum_{j} \left( i g_1 \hat{c}_{\uparrow, j+1}^{\dag} \hat{c}_{\downarrow, j} + {\rm h.c.}\right),
    \end{aligned}
\end{equation}
with the subscript indicating the site and spin. As mentioned in Sec.~\ref{sec:prelims}, for fermions, the dynamical matrix and Hamiltonian are one and the same. In momentum space, the BdG hamiltonian is,
\begin{equation}
    H_{\rm SHN}(k) = w \cos k + i \Delta \sin k \sigma_z + g_1 \sin k \sigma_x.
\end{equation}
This is known to host an SPSE of the form Eq.~\eqref{eq:TRS_def}, with symmetry operator $T = i \sigma_y$, and 
\begin{equation}\label{eq:SHN_kTRS}
    i \sigma_y H_{\rm SHN}(k)^T (-i \sigma_y) = H_{\rm SHN}(-k).
\end{equation}
This symmetry is formally a TRS, but also has a clear physical interpretation as a TRS, since it is the symmetry of coupling a Hatano-Nelson model to its time-reversed partner.

The strong analogy between the $g_1$-BKC and the SHN can be made concrete by observing that 
\begin{equation}\label{eq:SHNg1_sim}
    H_{\rm SHN}(k) = P M_{g_1}(k - \pi/2) P^{\dag}, \qquad P := \begin{pmatrix}
        1 & 0 \\ 0 & i
    \end{pmatrix},
\end{equation}
hence the two dynamical matrices are unitarily equivalent up to a $\pi/2$ shift in momentum space. This acts on $\sigma_y$ as,
\begin{equation}\label{eq:Py_invariance}
    P \sigma_y P = i \sigma_y.
\end{equation}

Putting Eqs.~(\ref{eq:SHN_kTRS}, \ref{eq:SHNg1_sim}, \ref{eq:Py_invariance}) together, we have 
\begin{equation}
    \begin{aligned}
        &i \sigma_y H_{\rm SHN}(k)^T (-i \sigma_y) = H_{\rm SHN}(-k) \\
        &\iff i \sigma_y P^{\dag} M_{g_1}(k-\pi/2)^T P (-i \sigma_y) \\
        &\qquad \qquad = P M_{g_1}(\pi/2 - k) P^{\dag} \\
        &\iff  \sigma_y M_{g_1}(k-\pi/2)^T  \sigma_y = M_{g_1}(\pi/2 - k),
    \end{aligned}
\end{equation}
which we see is simply Eq.~\eqref{eq:BKC_SPSE}. What is understood to be a formal consequence of sublattice symmetry in a bosonic model is actually a time-reversal symmetry in a fermionic model! 

To conclude, this discussion demonstrates that the topological origins of the skin effect in the $g_1$-BKC and the SHN are in fact one and the same. The $g_1$-BKC enjoys the benefit of arising without any post-selection, and in this light may serve as an effective SHN. The SHN has been extensively studied in the non-Hermitian community \cite{Kawabata_PRB_2020_symplectic_class, Kawabata_PRX_2019_symmetry_topology_NH, Okuma_PRL_2020_topological_origin, Kawabata_PRX_2023_EPT_from_NHSE}, which leaves us with the question -- which of its features survive in this effective setting?

\subsection{BKC-like models and the general 38-fold way non-Hermitian classification}\label{ssec:BKC-38}

There is a systematic topological classification of non-Hermitian
Hamiltonians due to Kawabata et al. \cite{Kawabata_PRX_2019_symmetry_topology_NH}. To derive the topological invariant \eqref{eq:doubled_winding}, we relied solely on the presence of time-reversal symmetry (TRS). A finer classification can, however, be obtained by taking the remaining symmetries into account \footnote{We emphasize that a given model may belong to different symmetry classes, since one is always free to disregard symmetries that are not relevant to the problem under consideration.}. While this refined classification is useful in general, it is not the most convenient framework for the analysis carried out in this work. Nevertheless, for the benefit of interested readers, we provide in this section a dictionary between the classification of Kawabata et al. and the formalism adopted in our work.

To distinguish their notation from ours, the terminology of Kawabata et
al. is written in an italic font. In their language, the generators of the
different classes are the internal symmetries \emph{TRS}, \emph{PHS},
\emph{TRS}$^{\dagger}$ and \emph{PHS}$^{\dagger}$. Both \emph{TRS} and
\emph{PHS} split into two variants.  This doubling is a consequence of the non-Hermiticity of the Hamiltonian (or dynamical) matrices involved (i.e.~complex conjugation and transpose operations are no longer equivalent).  The chiral symmetry likewise splits in two, namely \emph{CS} and \emph{SLS}. In addition, there may be a further internal symmetry known
as pseudo-Hermiticity, \emph{pH}. 

The symmetries identified by Kawabata et al are not all independent, but instead obey the following equivalence relations:
\begin{align}
CS & =TRS\times PHS\nonumber \\
 & =TRS^{\dagger}\times PHS^{\dagger},\nonumber \\
SLS & =TRS^{\dagger}\times PHS\nonumber \\
 & =TRS\times PHS^{\dagger},\\
TRS & =pH\times TRS^{\dagger},\nonumber \\
PHS & =pH\times PHS^{\dagger},\nonumber \\
SLS & =pH\times CS.\nonumber 
\end{align}
Depending on the phase convention, a phase factor may appear in these
relations; the equalities should therefore be understood as holding up
to a phase factor.

Table~\ref{tab:Symmetry_summary} summarizes these symmetries and how
they are realized in the context of the BKC. 
For simplicity, we omit qPHS from this table. The presence of qPHS yields an additional  \emph{PHS}$^{\dagger}$, i.e.~a symmetry that is independent of the aPHS symmetry built into the system. 
The $TRS$, $PHS^{\dagger}$ and $CS$
symmetries, which were not introduced elsewhere in the manuscript, were
computed using the relations ${\cal T}_{-}=\eta^{-1}{\cal C}_{-}$,
${\cal T}_{+}=\eta{\cal C}_{+}^{-1}$ and $\Gamma=\eta S$.

\begin{table*}[th!]
\begin{centering}
\begin{tabular}{|c|c|c|c|c|}
\hline 
\begin{cellvarwidth}[t]
\centering
Symmetry in

our notation
\end{cellvarwidth} & \begin{cellvarwidth}[t]
\centering
Symmetry in

Kawabata et al.
\end{cellvarwidth} & \begin{cellvarwidth}[t]
\centering
Action in

momentum space
\end{cellvarwidth} & \begin{cellvarwidth}[t]
\centering
Explicit expression

for BKC-like models
\end{cellvarwidth} & Square\tabularnewline
\hline 
 & \emph{TRS} & ${\cal T}_{+}{\cal M}^{*}\left(k\right){\cal T}_{+}^{-1}={\cal M}\left(-k\right)$ & $\sigma_{x}\otimes\sigma_{x}$ & ${\cal T}_{+}{\cal T}_{+}^{*}=1$\tabularnewline
\hline 
TRS & \emph{TRS}$^{\dagger}$ & ${\cal C}_{+}{\cal M}^{T}\left(k\right){\cal C}_{+}^{-1}={\cal M}\left(-k\right)$ & $\sigma_{x}\otimes\sigma_{y}$ & ${\cal C}_{+}{\cal C}_{+}^{*}=-1$\tabularnewline
\hline 
aPHS & \emph{PHS} & ${\cal C}_{-}{\cal M}^{T}\left(k\right){\cal C}_{-}^{-1}=-{\cal M}\left(-k\right)$ & $\mathbb{I}\otimes\sigma_{y}$ & ${\cal C}_{-}{\cal C}_{-}^{*}=-1$\tabularnewline
\hline 
 & \emph{PHS}$^{\dagger}$ & ${\cal T}_{-}{\cal M}^{*}\left(k\right){\cal T}_{-}^{-1}=-{\cal M}\left(-k\right)$ & $\mathbb{I}\otimes\sigma_{x}$ & ${\cal T}_{-}{\cal T}_{-}^{*}=1$\tabularnewline
\hline 
 & \emph{CS} & $\Gamma{\cal M}^{\dagger}\left(k\right)\Gamma^{-1}=-{\cal M}\left(k\right),$ & $\sigma_{x}\otimes\sigma_{z}$ & \tabularnewline
\hline 
SLS & \emph{SLS} & ${\cal S}{\cal M}\left(k\right){\cal S}^{-1}=-{\cal M}\left(k\right)$ & $\sigma_{x}\otimes\mathbb{I}$ & \tabularnewline
\hline 
pH & \emph{pH} & $\eta{\cal M}^{\dagger}\left(k\right)\eta^{-1}={\cal M}\left(k\right)$ & $\mathbb{I}\otimes\sigma_{z}$ & \tabularnewline
\hline 
\end{tabular}
\par\end{centering}
\caption{Summary of the different non-Hermitian symmetries. The
notation of Ref.~\cite{Kawabata_PRX_2019_symmetry_topology_NH} is indicated in an italic font. The action in
momentum space provides a general definition of each symmetry, and we
give the corresponding explicit expression for the cases considered in
this paper. Here $\cal{T}_\pm$, $\cal{C}_\pm$ designate the unitary part of the symmetries. For the explicit expressions, the first term in the tensor product acts on the folded zone indices $k$ and $k+\pi$ while the second term in the tensor product acts on the particle and hole degrees of freedom of the dynamical matrix. Here, $PHS^{\dagger}$ is generated from aPHS by applying the
pH symmetry; qPHS, by contrast, introduces a further $PHS^{\dagger}$
symmetry, which we discuss separately.}\label{tab:Symmetry_summary}
\end{table*}
\par
Classifying the different non-Hermitian models requires, beyond the
symmetries themselves, knowledge of the commutation relations between
the unitary and antiunitary symmetries. A direct computation gives
\begin{center}
\begin{tabular}{c|cccc}
 & ${\cal T}_{+}$ & ${\cal T}_{-}$ & ${\cal C}_{+}$ & ${\cal C}_{-}$\tabularnewline
\hline 
$\eta$ & $-1$ & $-1$ & $-1$ & $-1$\tabularnewline
${\cal S}$ & $+1$ & $+1$ & $+1$ & $+1$\tabularnewline
$\Gamma$ & $-1$ & $-1$ & $-1$ & $-1$\tabularnewline
\end{tabular}
\par\end{center}

With these connections established, we can now classify our BKC-like models using the framework of Kawabata et al. \cite{Kawabata_PRX_2019_symmetry_topology_NH}.  As discussed, the symmetries built into the model are aPHS and pH. Since $\eta$
anticommutes with ${\cal C}_{-}$, in the absence of any further
symmetry the BKC belongs to the class C+$\eta_{-}$.  
If we now have a BKC-like model that also has sublattice symmetry $SLS$, an alternate classification is possible.  Combining $SLS$ with aPHS (\emph{PHS}$^{\dagger}$) automatically yields $TRS^{\dagger}$, and applying $pH$ then generates $TRS$, $PHS$ and $CS$. In other words, $SLS$ endows the model with {\it all} the remaining symmetries at once. Since $\eta$ anticommutes with both $TRS$ and $PHS$, the resulting symmetry class is CI+$\eta_{--}$.

Finally, consider the case in which qPHS is present. We denote the
associated additional $PHS^{\dagger}$ symmetry (i.e.~distinct from the built-in particle hole symmetry) by ${\cal T}'_{-}$.  In our case, the unitary part of the symmetry is just the identity:
\begin{equation}
    {\cal T}'_{-}=\mathbb{I}\otimes\mathbb{I}.
\end{equation}
Combining ${\cal T}_{-}$, obtained from the product of aPHS and pH,
with ${\cal T}'_{-}$ produces a genuine unitary symmetry,
\begin{equation}
U={\cal T}_{-}^{'}{\cal T}_{-}=\mathbb{I}\otimes\sigma_{x},
\end{equation}
and hence the conserved quantity $\sigma_{x}$. This offers another way
to see that the system block-diagonalizes into the different quadrature
sectors, within which the symmetry analysis must now be carried out.
The surviving symmetries are those that commute with $U$, namely
$PHS^{\dagger}$, $SLS$ and $TRS$. Here $PHS^{\dagger}$ acts as plain
conjugation, ${\cal S}=\sigma_{x}$, and $TRS$ is the product of
$PHS^{\dagger}$ and $SLS$. Since $SLS$ commutes with $TRS$, the model
now sits in the class AI+${\cal S}_{+}$.


\section{Symmetry and stability}
\label{sec:symmetry_and_stability}

In this section, we discuss the relevance of the various symmetries evoked in the previous section for the stability analysis of the OBC system. In other words, we ask whether the eigenvalues of the dynamical matrix are real or not. As stated in the introduction, contrary to the fermionic case, stability plays an important physical role for bosonic systems as it will determine whether the particle number will diverge. An important fine point is that stability is not always concomitant with NHSE, i.e. the system can present an NHSE while still being unstable for OBC.

Our approach examines the effect of a given perturbation on an unperturbed Hamiltonian that respects sublattice symmetry (SLS). The central observation is that SLS forces the eigenvectors of the dynamical matrix to come in pairs of opposite energy. Any perturbation that couples such a pair can therefore hybridize the two modes and drive their energies together at zero, producing a zero-energy mode that destabilizes the system. In this sense the bare bosonic Kitaev chain (BKC)---like any bosonic system endowed with SLS---sits permanently on the threshold of dynamical instability. Remarkably however, for OBC, if the perturbation \emph{itself} respects SLS, one can show analytically that the matrix element responsible for coupling the pair vanishes identically. SLS thus plays a dual role: it poises the system at the edge of instability while simultaneously protecting it from crossing that edge. For PBC, there are additional degeneracies preventing this protection from happening and the system is generically unstable as we will see below.

\subsection{Perturbation in the pairing term. \label{subsec:perturbation in pairing}}

We begin our discussion with the case where the pairing term is treated as the perturbation. Owing to the original bosonic
algebra and the Hermiticity of the second quantized Hamiltonian, $M^{(0)}$
can be, without loss of generality, written in the position basis as: 
\begin{equation}
M^{(0)}=\begin{pmatrix}m & 0\\
0 & -m^{T}
\end{pmatrix},
\end{equation}
where $m$ is an Hermitian $N\times N$ matrix with real eigenvalues. $m$ corresponds to the ``particle'' sector and $-m^T$ to the ``hole'' sector.
Let $\left|\varepsilon_{k}\right\rangle $ be an eigenvector of $m$,
$m\left|\varepsilon_{k}\right\rangle =\varepsilon_{k}\left|\varepsilon_{k}\right\rangle $.
We require that $M^{\left(0\right)}$ obey the SLS,  $SM^{\left(0\right)}S^{-1}=-M^{\left(0\right)}$. We further assume that, for OBC, the spectrum $\varepsilon_k$ is non-degenerate--an assumption we verify for every model considered in this work. Note, however, that we do not assume translational invariance in this section.

Let $\left|\varepsilon_{k}\right\rangle =\sum_{j}u_{k,j}\left|j\right\rangle $
be the decomposition of $\left|\varepsilon_{k}\right\rangle $ in
the position basis. An eigenvector of $m^{T}$ is obtained by applying
the complex conjugation operation in position basis (recall that the
complex conjugation operation applied to a vector is basis dependent).
Let us denote this operation by $\left|\overline{\varepsilon_{k}}\right\rangle :=\sum_{j}u_{k,j}^{*}\left|j\right\rangle $.
We have

\begin{align}
-m^{T}\left|\overline{\varepsilon_{k}}\right\rangle  & =-\varepsilon_{k}\left|\overline{\varepsilon_{k}}\right\rangle .
\end{align}
Let $\left|0\right\rangle $ be the null vector of size $N$. The
eigenvectors of $M^{(0)}$ are then obtained as $\left|\varepsilon_{k},\bullet\right\rangle :=\left|\varepsilon_{k}\right\rangle \oplus\left|0\right\rangle $,
$\left|-\varepsilon_{k},\circ\right\rangle :=\left|0\right\rangle \oplus\left|\overline{\varepsilon_{k}}\right\rangle $. Note that this is a consequence of the aPHS (Eq.~\eqref{eq:aPHS}), where we have explicitly identified the particles and the holes.

We now treat as a perturbation the part of the Hamiltonian that comprises
pair creation or annihilation terms. Denote, in position space, the
perturbation by 
\begin{equation}
M^{(1)}:=\lambda\begin{pmatrix}0 & V\\
-V^{*} & 0
\end{pmatrix}
\end{equation}
where $V$ is a symmetric $N\times N$ matrix and $\lambda$ a control
parameter that we will take to be small. Again, the structure of the
matrix is imposed from the bosonic algebra and Hermiticity of the
original Hamiltonian. Because $M^{(0)}$ is block diagonal and $M^{(1)}$
is block off-diagonal, the first order correction in $\lambda$ to
the spectrum must come from the degenerate eigenspaces.

As we supposed that the spectrum of $M^{\left(0\right)}$ is non-degenerate for OBC, the only way to form a degenerate subspace is with a
pair $\left\{ \left|\varepsilon_{k},\bullet\right\rangle ,\left|-\varepsilon_{k'},\circ\right\rangle \right\} $
for which $\varepsilon_{k}=-\varepsilon_{k'}$. Another way to phrase
this is that one needs to form a pair of modes with $0$ energy in
order to fulfill a resonance condition for the modes $k$ and $k'$.

Now if the OBC system possesses a SLS $S$, then for
each eigenvector $\left|-\varepsilon_{k},\circ\right\rangle $, there
is a corresponding opposite eigenvector $\left|\varepsilon_{k},\circ\right\rangle =S\left|-\varepsilon_{k},\circ\right\rangle $.
The associated ``hole'' eigenvector $\left|\varepsilon_{k},\circ\right\rangle $
then has eigenvalue $\varepsilon_{k}$ and the pair $\left\{ \left|\varepsilon_{k},\bullet\right\rangle ,\left|\varepsilon_{k},\circ\right\rangle \right\} $
provides the desired degenerate subspace. Thus, as stated previously, SLS brings the system to the edge of instability as any hybridization of the two modes will provoke a dynamical instability. However, if the perturbation itself fulfills SLS, the system is perturbatively protected from it and remains stable, as we will now see.

To first order in $\lambda$, the corrections to the eigenvalue $\varepsilon_{k}$,
$\delta\varepsilon_{k}^{\pm}$ are given by the eigenvalues of the
$2\times2$ matrix which comprises the elements of $M^{(1)}$ restricted
to the degenerate subspace that we call $M_{\varepsilon_{k}}^{\left(1\right)}$:
\begin{align}
M_{\varepsilon_{k}}^{\left(1\right)} & :=\lambda\begin{pmatrix}0 & a\\
-a^{*} & 0
\end{pmatrix},\\
\delta\varepsilon_{k}^{\pm} & =\pm i\lambda\left|a\right|^{2},
\end{align}
with $a=\left\langle \varepsilon_{k}\right|VS\left|\overline{\varepsilon_{k}}\right\rangle $
and thus one generically expects that any infinitesial perturbation
will lead to instabilities. However, in the case where $M^{(1)}$
also fulfills SLS $SVS^{-1}=-V$, we have 
\begin{align}
a & =\left\langle \varepsilon_{k}\left|VS\right|\overline{\varepsilon_{k}}\right\rangle =-\left\langle \varepsilon_{k}\left|SV\right|\overline{\varepsilon_{k}}\right\rangle ,\nonumber \\
 & =-\left(\left\langle \overline{\varepsilon_{k}}\left|V^{\dagger}S^{\dagger}\right|\varepsilon_{k}\right\rangle \right)^{*}=-\left\langle \varepsilon_{k}\left|VS\right|\overline{\varepsilon_{k}}\right\rangle ,\label{eq:argument_stability}\\
 & =0.\nonumber 
\end{align}
where we also took advantage of the fact that $V^{T}=V$, $S^{T}=S$ and $\left(\left\langle \phi\right|A\left|\psi\right\rangle \right)^{*}=\left\langle \psi\right|A^{\dagger}\left|\phi\right\rangle =\left\langle \overline{\phi}\right|A^{*}\left|\overline{\psi}\right\rangle $.
We thus see that in the presence of SLS, the pair
creation and annihilation term do not lift the degeneracy and provoke
an instability. 

Note that this argument \emph{does not }apply to the PBC case. The
hypothesis that breaks down is the fact that the spectrum is in this
case degenerate. For the sake of this argument, we assume that the system is translationally invariant, with eigenvectors indexed by momentum $k$. There are then two ways to create a $0$ energy pair,
either by coupling modes with opposite momenta $\left\{ k,-k\right\}$ that
are related by complex conjugation or by coupling modes that are related
by SLS with momenta $\left\{ k,k\pm\pi\right\} $
. In the OBC case, these two couples are the same which leads to the
argument of cancellation of Eq.~\eqref{eq:argument_stability}. In
the PBC case, this is not the case anymore and this argument is no
longer valid. The bare BKC is an example of this situation where an
infinitesimal $\Delta$ leads to instabilities.
\subsection{Perturbation in the hopping term}

We now discuss the case where the bare Hamiltonian
is $\hat{H}_{{\rm BKC}}$ and the perturbation is of the form 
\begin{equation}
\hat{H}_{d}=\frac{1}{2}\sum_{j}\left(g_{d}\hat{a}_{j+d}^{\dag}\hat{a}_{j}+{\rm H.c.}\right),
\end{equation}
with $g_{d}\in\mathbb{C}$ and denote $M_{d}$ the associated dynamical
matrix. Such a perturbation breaks the sublattice symmetry when $d$
is even, and preserves it otherwise. 

We start by providing some comments on the stability
of the BKC against such perturbations. This is a priori outside of
the scope of the perturbative analysis discussed in the previous section
as the bare Hamiltonian comprises pair creation terms while the perturbation
is particle number conserving.

However, as is well-known \cite{McDonald_PRX_2018_BKC}, the following canonical transformation
maps $\hat{H}_{{\rm BKC}}$ onto a simple tight-binding chain: 
\begin{align}
\begin{pmatrix}\hat{a}_{j}\\
\hat{a}_{j}^{\dagger}
\end{pmatrix} & =U\begin{pmatrix}\hat{b}_{j}\\
\hat{b}_{j}^{\dagger}
\end{pmatrix},\label{eq:rotation}
\end{align}
with $U:=i^{j}\begin{pmatrix}\cosh\left(rj\right) & -\sinh\left(rj\right)\\
-\sinh\left(rj\right) & \cosh\left(rj\right)
\end{pmatrix}$ leads to 
\begin{equation}
\hat{H}_{{\rm BKC}}=\frac{1}{2}J\sum_{j}\left(\hat{b}_{j}^{\dagger}\hat{b}_{j+1}+{\rm H.c.}\right).
\end{equation}
The transformation (\ref{eq:rotation}) induces in turn a symplectic
rotation on the dynamical matrices
\begin{align}
M'_{{\rm BKC}} & :=U^{-1}M_{{\rm BKC}}U,\\
M'_{d} & :=U^{-1}M_{d}U.
\end{align}
As $M'_{d}$ will not be in general block-diagonal anymore, we
split it into a diagonal and an off-diagonal part: 
\begin{equation}
M'_{d}=:M_{d}^{\prime\text{ diag}}+M_{d}^{\prime\text{ off-diag}}.
\end{equation}
In the framework of the perturbative theory previously introduced,
the diagonal part of the latter equation can be reabsorbed in the
bare non-perturbed Hamiltonian and the off-diagonal part treated as
the perturbation: 
\begin{align}
M^{(0)} & :=M'_{{\rm BKC}}+M_{d}^{\prime \text{ diag}},\\
M^{(1)} & :=M_{d}^{\prime\text{ off-diag}}.
\end{align}
As the sublattice symmetry is preserved under the action of $U$,
the previous perturbative analysis for the stability applies to this
case.

Finally, we note that similar arguments cannot be made using the qPHS. In fact as we will see in the next section, the BKC appears to be highly unstable against generic qPHS-preserving but sublattice symmetry-breaking perturbations.


\bigskip

\section{Generalized BKC physics via symmetry-enriched bosonic pairing models }\label{sec:classification}

\subsection{Couplings and symmetry}

Having established in Sec.~\ref{sec:symmetry_blitz} the two key symmetries underlying bare BKC's skin effect,
we now turn to more general models, namely the class of general 1D Hermitian bosonic pairing models given in Eq.~\eqref{eq:gen_TI_QBH}.  Our goal is to classify these in terms of the two symmetries, and the kinds of skin effects that can be supported.  
Models of the form of Eq.~\eqref{eq:gen_TI_QBH} can be classified in terms of (A) whether or not they have quadrature decoupling (i.e.~qPHS), and (B) whether or not they have sublattice symmetry. This classification lets us understand whether a particular model can support a skin effect 
whose topological origin matches that of the BKC's skin effect.  Hence, it lets us understand which of these models generalizes the BKC
\footnote{
We note that the absence of either symmetry does not indicate the absence of a skin effect, as there remain less well-understood effects, such as the reentrant skin effect \cite{Yokomizo_PRB_2021_nonbloch_bosonic}, which appear to defy topological classification. These fall outside the scope of the current work.}.
This results of the analysis we present here is summarized in  Fig.~\ref{table:symmetric_schematic}.

\textit{Quadrature decoupling (qPHS) -- } As noted in the discussion after Eq.~\eqref{eq:qPHS_equiv}, a dynamical matrix is quadrature decoupled if it is diagonal in some $k$-independent basis, which for the basic BKC is the $\sigma_x$ basis, and corresponds formally to the qPHS particle-hole symmetry. Alternately, models of the form of Eq.~\eqref{eq:AB_parameterization} have qPHS if there exists a local $U(1)$ gauge transformation such that the Hamiltonian in the creation and annihilation basis becomes purely imaginary. On the level of terms in Eq.~\eqref{eq:AB_parameterization}, this translates to models with $g_d = 0$ for all $d$, and the quantity $\arctan(\Delta_d/\eta_d)$ being independent of $d$ (this allows us to rotate the phase in the squeezing term away with a $d$-independent rotation). In the examples of this section, we will consider concrete models that extend the basic BKC (i.e. build on top of a base model that already has $w_1, \Delta_1$ terms). In that case, the base $\Delta_1$ term already selects the preferred set of quadratures, so for qPHS, we require that all $g_d, \eta_d$ terms are $0$.


\textit{Sublattice symmetry -- } Inspecting Eq.~\eqref{eq:sublattice_req} immediately tells us that any non-zero coupling $g_d, \eta_d$ with $d$ even will break sublattice symmetry, whereas any model with only odd-$d$ couplings non-zero retains sublattice symmetry. Such models then host a symmetry-protected $\mathbb{Z}_2$ topological invariant, which can then potentially yield a symmetry-protected NHSE.  

With these simple considerations in mind, we can now study the phenomenology of particular models in each row of Table.~\ref{table:symmetric_schematic}. Following the results of Sec.~\ref{sec:symmetry_and_stability}, we will also discuss the stability of these models. As mentioned, in this section we only consider `BKC-like' models, in the sense they are all built by adding couplings to $\hat{H}_{\rm BKC}$, but we note that the ideas hold generally for translationally-invariant QBHs. In this section, we restrict to quadratic models, but we provide some comments on how this phenomenology may extend to interacting models in App.~\ref{sec:interactions}.

\begin{figure}
\centering{}\includegraphics[width=\columnwidth]{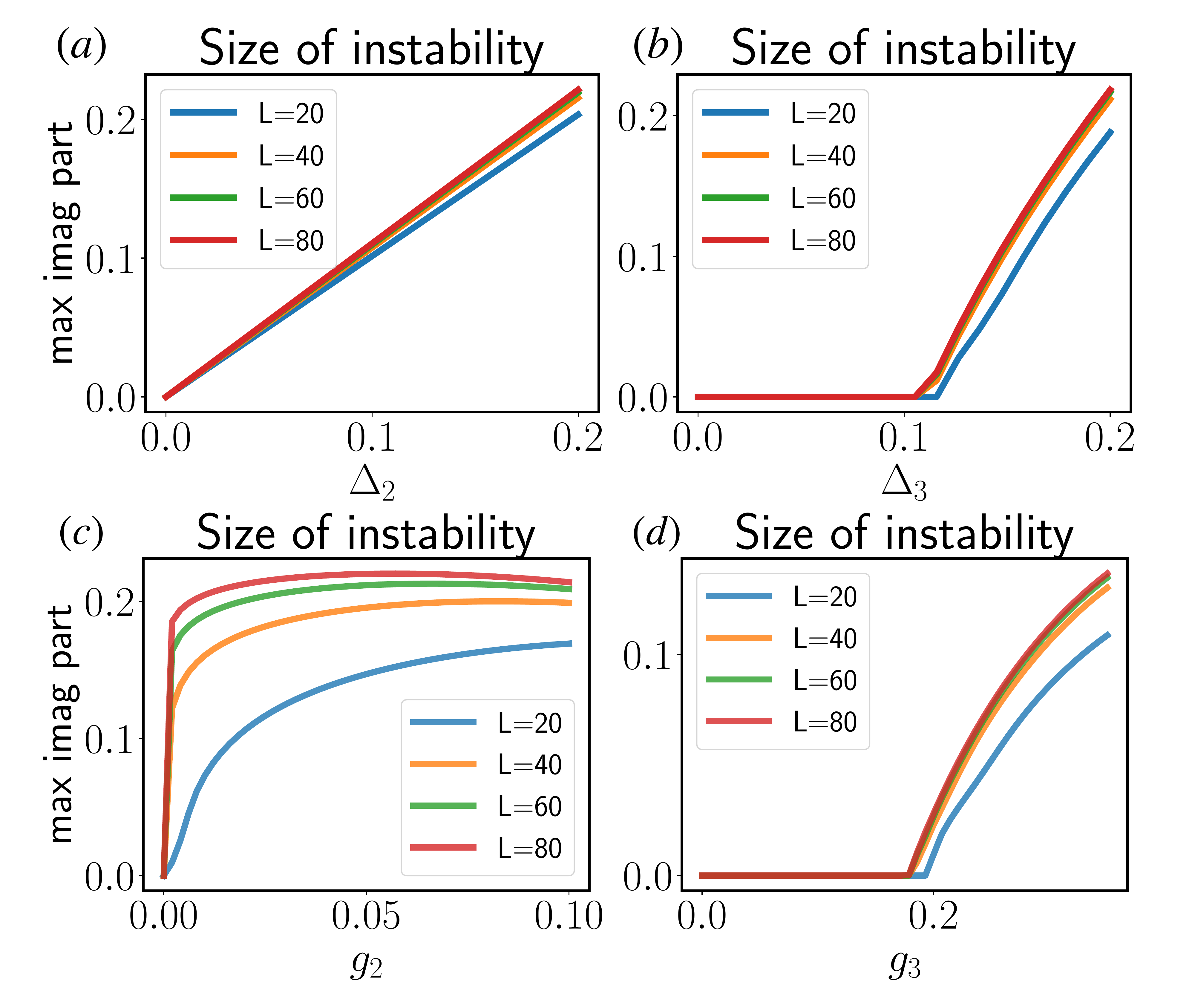}\caption{Plot of the maximum imaginary part of any eigenvalue of (a) $\Delta_2$-BKC, (b) $\Delta_3$-BKC, against the value of $\Delta_{2, 3}$, and (c) $g_2$-BKC, (d) $g_3$-BKC, against the value of $g_{1, 2}$. The other (underlying bare BKC) parameters are set to $w=1, \Delta = 0.25$, and all values are computed under OBC for a range of system sizes $L = 20, 40, 60, 80$. The perturbations depicted in the left column break sublattice symmetry; accordingly the BKC under OBC is immediately unstable against such perturbations. Conversely, the perturbations depicted in the right column preserve sublattice symmetry; accordingly the stability of BKC under OBC is protected against such perturbations.}\label{fig:instab_combined}
\end{figure}


\subsection{Sublattice symmetry without quadrature decoupling\label{subsec:with_g3}}

We first consider a model we term the $g_3$-BKC, which serves as a paradigmatic example of a model that has sublattice symmetry {\it without} any qPHS (i.e.~there is no quadrature decoupling).  We are thus breaking one of the two symmetries of the bare BKC.  
\begin{figure*}
\centering{}\includegraphics[width=\linewidth]{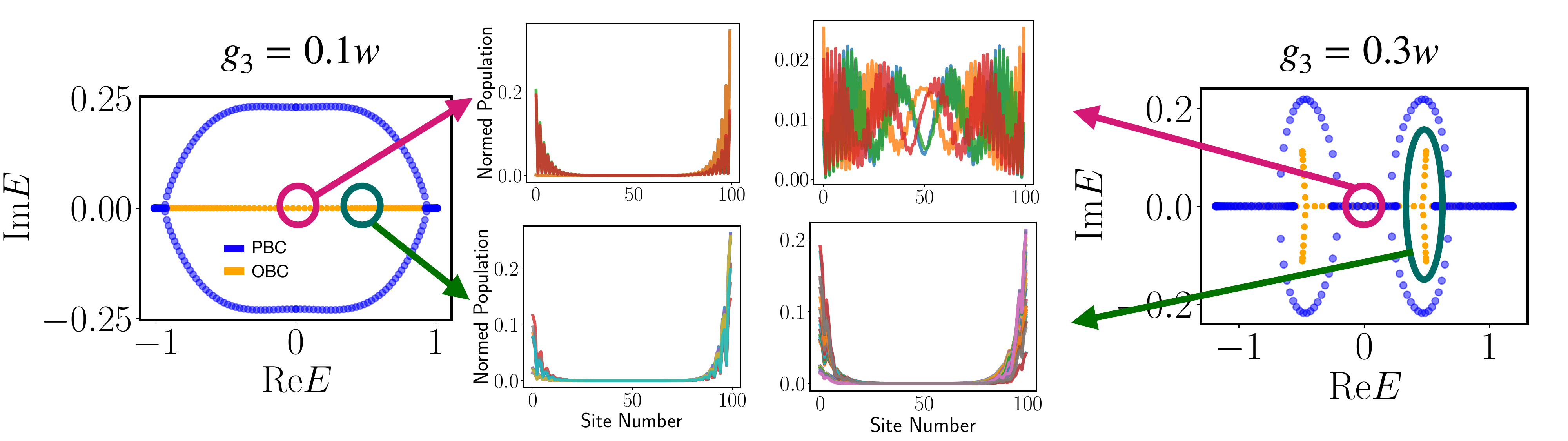}\caption{Plot depicting the localization-delocalization transition (or lack thereof) of particular sets of eigenstates in the spectrum of the $g_3$-BKC. The left-most figure depicts the spectrum of the $g_3$-BKC for $w = 1, \Delta = 0.25w, g = 0.1w$, with the OBC spectrum in orange and the PBC spectrum in blue. In this case, both states around $E=0$ and $E = \pm w \sin (\pi/6)$ are localized; in the second column of plots we plot a selection of such states. These are the states corresponding to eigenvalues whose real part is less than $0.05w$ away from $E = 0 (E = w\sin \pi/6)$. The quantity being plotted is the normalized population supported on each lattice site. The left-most figure depicts the spectrum of the $g_3$-BKC for $w = 1, \Delta = 0.25w, g_3 = 0.3w$, with the OBC spectrum in orange and the PBC spectrum in blue. In this case, states around $E=0$ are no longer localized, as seen in the sample of states in the third column, whereas states around $E=w\sin \pi/6$ remain localized.}\label{fig:g3-spectrum}
\end{figure*}

The $g_3$-BKC model is defined by:
\begin{equation}\label{eq:g3_BKC}    \hat{H}_{g_3} = \hat{H}_{\rm BKC} + \frac{1}{2} \sum_{j} \left( g_3 \hat{a}_{j+3}^{\dag} \hat{a}_j + h.c. \right).
\end{equation}
As discussed above, the $g_3$ 
term adds a term with a real coefficient to the bare BKC model, and as such will necessarily have dynamics that couples the canonical $q,p$ quadratures.  However, sublattice symmetry is preserved, as there are no even distance $d$ couplings.  Following the analysis in Sec.~\ref{sec:symmetry_blitz}, sublattice symmetry immediately implies that this model's dynamical matrix has an effective TRS, something that lets us define a meaningful topological invariant and enables a symmetry-protected skin effect (SPSE).  Crucially, the $g_3$-BKC does not have physical TRS in the form of a local anti-unitary symmetry of the second quantized Hamiltonian, something we discussed explicitly in Sec.~\ref{lsubsec:TRSNotTRS}.



We first consider the dynamical stability of the $g_3$-BKC with open boundary conditions. 
Recall the general argument of Sec.~\ref{sec:symmetry_and_stability}:
chiral symmetry leads to a surprising perturbative robustness against pairing-induced dynamical instabilities, even though the unperturbed $\Delta=0$ model contains zero-energy pairs. Importantly, this prediction is only perturbative: it guarantees stability against infinitesimal pairing terms, but does not imply stability over a finite parameter range. 

As a probe of stability, we numerically diagonalize the OBC dynamical matrix of the $g_3$-BKC, and plot in Fig.~\ref{fig:instab_combined}d the largest imaginary part of any eigenvalue, with zero indicating dynamical stability. As shown in 
this figure
for several system sizes, the $g_3$-BKC appears to remain dynamically stable over a finite range of perturbation strengths. Surprisingly, while this validates our prediction, it also suggests a stronger protection against instability that goes beyond the perturbative regime (and the validity of our proof in Sec.~\ref{sec:symmetry_and_stability}). We leave the study of this finite-range stability to future work. The eventual onset of instability occurs at some $g^*<\Delta$, which, as we see shortly, does not correspond to a topological transition associated with the NHSE.

Next, we can study the NHSE in the $g_3$-BKC model. Since the model retains sublattice symmetry, Eq.~\eqref{eq:sublattice_winding} 
yields a meaningful, quantized spectral winding number that can yield a skin effect.
The (PBC) momentum-space dynamical matrix is,
\begin{equation}
    M(k) = w \sin(k) - i \Delta \cos (k) \sigma_x - g_3 \cos(3k) \sigma_z
\end{equation}
with eigenvalues,
\begin{equation}\label{eq:g3_spectrum}
    E_{\pm}(k) = w \sin k \pm  \sqrt{g^2_3 \cos^2(3k) -  \Delta^2 \cos^2 (k)}.
\end{equation}
The form of the spectrum provides useful intuition into how the $g_3$ term modifies the physics.  For $\Delta = 0$, this term adds a $\cos(3k)$ term to the spectrum, and hence gives a net energy to a pair of excitations with momenta $k$ and $-k$.  This detuning hence cuts off instability at weak $\Delta$ for most $k$, i.e.~it keeps the square-root in Eq.~\eqref{eq:g3_spectrum} real.  Crucially, this detuning effect vanishes at $k=\pm \pi/6$; for this momentum, we will have dynamical instability in the PBC system for arbitrarily small $\Delta$. 

We will study how the topology of our model changes as $g_3$ is varied with all other parameters held fixed.  To identify topology via our sublattice symmetry invariant, we need to identify reference energies (i.e.~specific point gaps) around which there might be non-trivial spectral winding.    
One natural choice is $E=0$ (or sufficiently close to it). In that case, $\nu_{\rm sl}(E) = 1$ when $g_3 < \Delta$, and $\nu_{\rm sl}(E) = 0$ when $g_3 > \Delta$, similar to the $g_1$-BKC. Given our discussion of the spectrum above, another choice of reference energy is given by the special role of $k= \pm \pi/6$, namely reference energies $E = \pm w \sin(\pi/6)$ (or sufficiently close). Unlike the choice $E=0$, for these non-zero reference energies, we find that $\nu_{\rm sl}(E) = 1$ for {\it all} values of $g_3$.  

We thus have a kind of behaviour that has no analogue in the bare BKC: for one base energy $E=0$, we have a transition from a non-trivial to trivial winding number as $g_3$ is increased, whereas for another base energy 
$E = \pm w \sin(\pi/6)$, the winding number remains nontrivial for all $g_3$.  The behaviour of these winding numbers immediately implies non-trivial skin effect physics in the OBC system.    
It implies that for OBC eigenstates with energies in the neighborhood of $E=0$, we will observe a topological transition; for $g_3 < \Delta$, these states are localized, but the localization is lost when $g_3 > \Delta$. Note that this feature is shared by the $g_d$ for every odd $d$. Alternatively, for OBC states near $E = \pm w \sin(\pi/6)$ are always localized, for any value of $g_3$. We explicitly plot such states in Fig.~\ref{fig:g3-spectrum}. Other values of odd $d$ exhibit similar phenomena, with special non-zero values of reference energies again determined by values of $k$ where the PBC system is unstable for infinitesimal $\Delta$. The behavior of this model highlights the difference between the features outlined in the introduction.  In the basic BKC, this is hard to see -- the regimes of $w, \Delta$ for which feature (2) holds are precisely the regimes for which feature (1) holds. In stark contrast, for the new $g_3$-BKC model these features no longer coincide-- increasing $g_3$ causes feature (2) to be lost even though the NHSE remains present. 

Finally, we provide some comments on potential applications of this model. First, recall that for these systems the OBC spectrum must be contained in the interior of the PBC spectrum \cite{Okuma_PRL_2020_topological_origin,Znidaric_PRR_2022_solvable_NHSE, Bhat_JPA_2025_boom_bust_pseudospectra}. It follows that OBC instability is only permissible for energies whose real parts have non-zero winding $\nu({\rm Re} E)=1$. Since the complex part of the PBC spectrum, Eq.~\eqref{eq:g3_spectrum}, narrows around $E = \pm w \sin (\pi/6)$, we conclude that when $g$ is large, the OBC spectrum may only exhibit instability for states whose real parts lie around $\pm w \sin (\pi/6)$, which interestingly, suggests a frequency-selected amplification in this model. In particular, modes with frequency close to $\pm w \sin (\pi/6)$ will be amplified much more strongly than any other part of the spectrum. Such a phenomenon may be useful in various physical applications, eg. sensing.

\subsection{Sublattice symmetry-broken but quadrature decoupled}

We next consider a different generalized model that keeps the qPHS (quadrature decoupling) symmetry of the bare BKC model, but breaks its sublattice symmetry.  A minimal model achieving this is the $\Delta_2$-BKC, defined by the Hamiltonian:
\begin{equation}\label{eq:H_Delta2}
    \hat{H}_{\Delta_2} = \hat{H}_{\rm BKC} +  i \Delta_2 \frac{1}{2} \sum_j  \left( \hat{a}_{j+2}^{\dag} \hat{a}_j^{\dag} - {\rm H.c.} \right).
\end{equation}
As the $\Delta_2$ term is an imaginary term in $\hat{H}$, it keeps the canonical $q,p$ quadratures decoupled.  However, as it is an even-range coupling, it breaks sublattice symmetry.  

We start again by considering dynamical stability.
To study stability of the OBC system, in Fig.~\ref{fig:instab_combined}a, we fix $w = 1, \Delta = 0.25$, and plot the largest imaginary eigenvalue of the dynamical matrix against the size of a $\Delta_2$ perturbation. We see that the BKC is clearly first-order unstable against such a sublattice-symmetry breaking perturbation: an arbitrarily small $\Delta_2$ yields imaginary eigenvalues. This should be contrasted against Fig.~\ref{fig:instab_combined}b, which is an analogous plot for a $\Delta_3$ perturbation to the bare-BKC model. Note that in contrast to the case of chemical potential (see Sec.~\ref{ssec:both_sym_broken}), these instabilities do not depend strongly on system size.

\begin{figure*}[t!]
\centering

\begin{minipage}[t]{0.47\textwidth}
\centering
\vspace{0pt}
\includegraphics[
  width=\linewidth,
  height=0.22\textheight,
  keepaspectratio
]{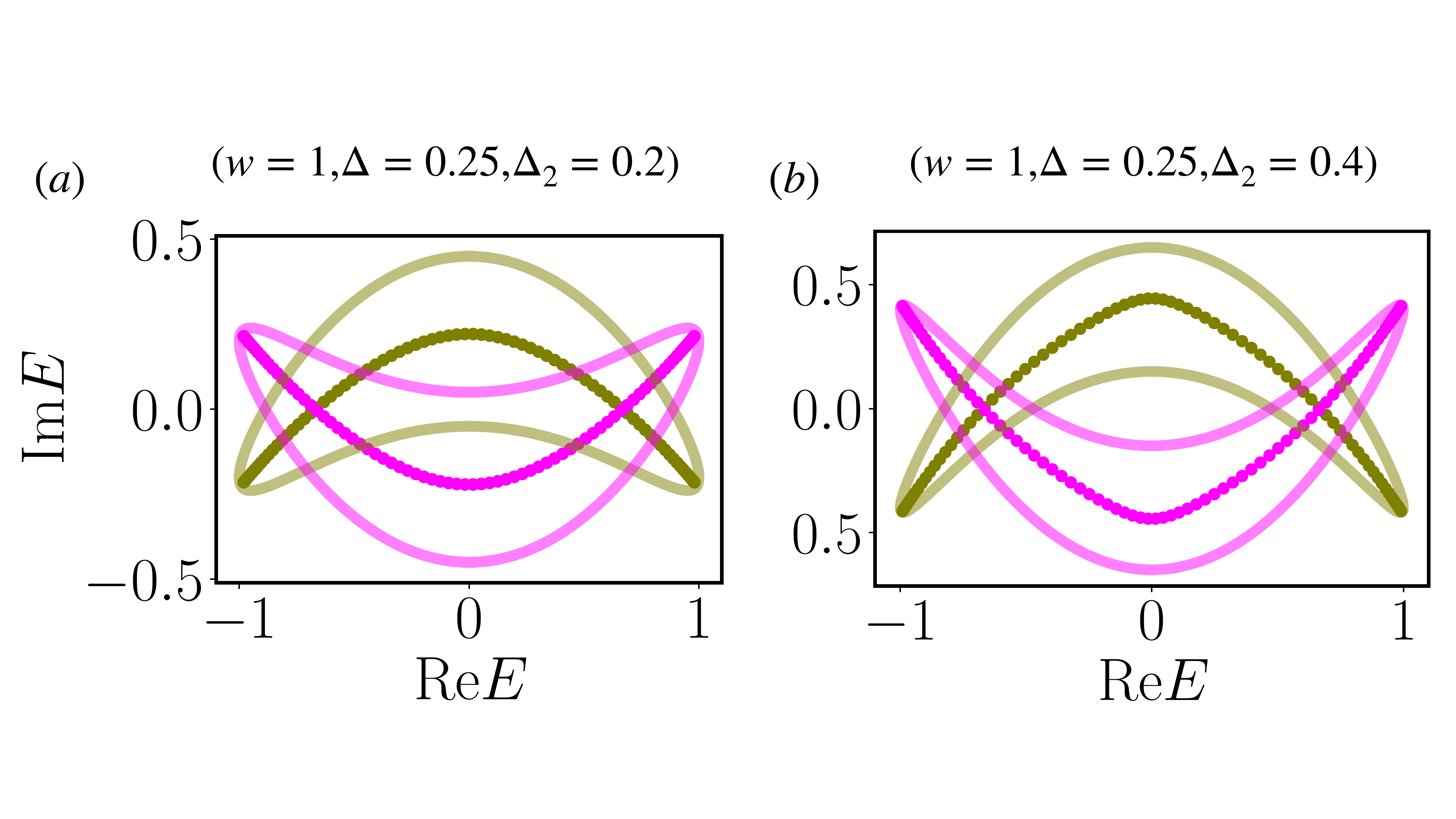}
\caption{Spectra for the $\Delta_2$-BKC (see Eq.~\eqref{eq:H_Delta2}). The olive curves depict the $q$ band, and the magenta curves depict the $p$ band. The solid points are the eigenvalues under OBC and the translucent lines are for PBC. In both (a), (b), the underlying BKC has parameters $w = 1, \Delta = 0.25w$. The plots (a), (b) differ only by the size of the parameter $\Delta_2 = 0.2w$ and $\Delta_2 = 0.4w$. In both panels, the PBC version of the $q$ ($p$) band winds around the entirety of the OBC version of the $q$($p$) band only, but not the other $p$ ($q$) band. In (a), both bands wind around $E=0$, whereas in (b), neither band does.}
\label{fig:spectrum_D2}
\end{minipage}
\hfill
\begin{minipage}[t]{0.47\textwidth}
\centering
\vspace{20pt}
\includegraphics[
  width=0.85\linewidth,
  height=0.15\textheight,
  keepaspectratio
]{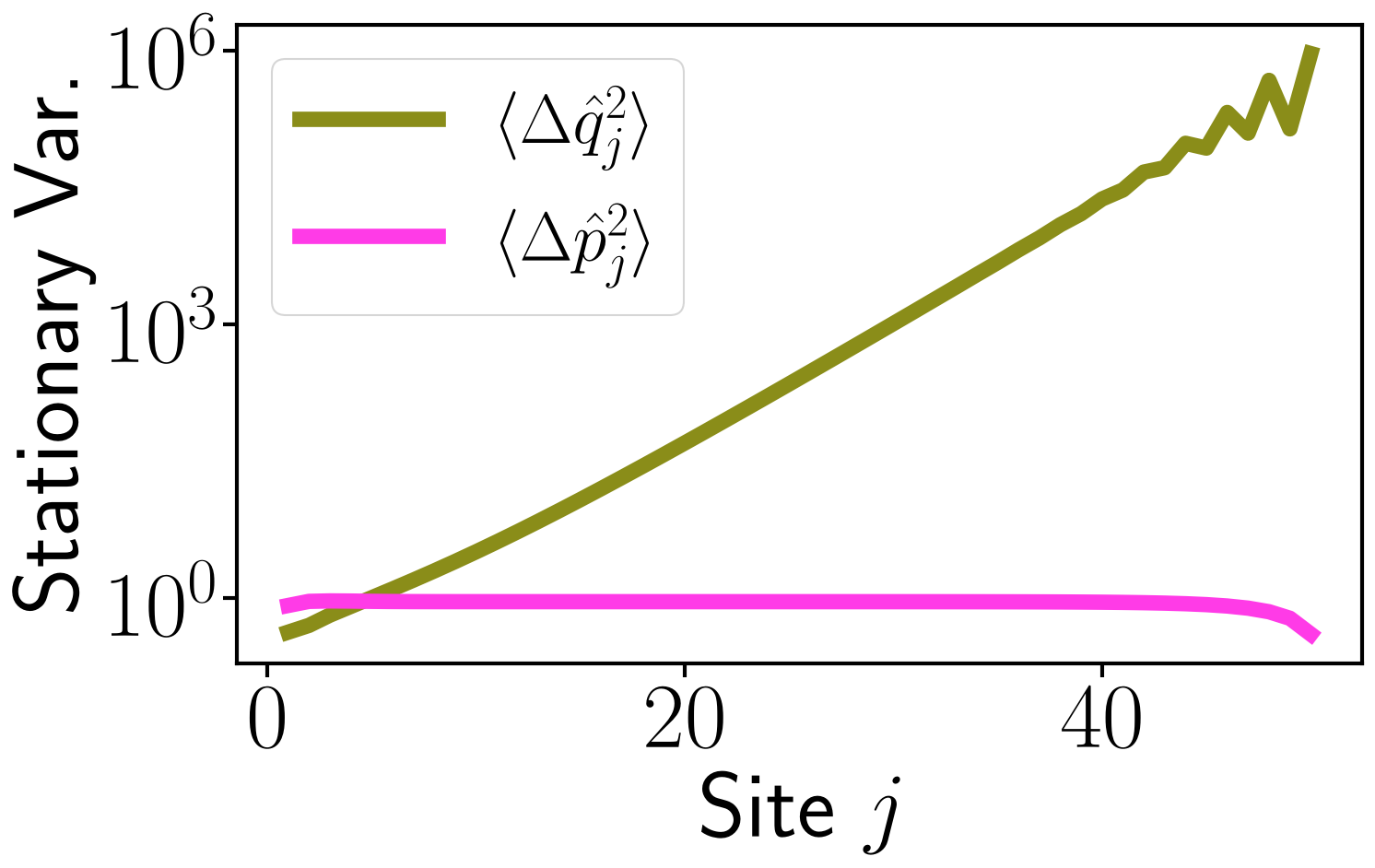}
\caption{Steady-state position-dependent quadrature variances of the $\Delta_2$-BKC subject to background loss, $w=1$, $\Delta=0.25$, $\Delta_2=0.2$, $L=50$. The effect of the loss is uniformly translate the spectra down by ${\rm Im E} \rightarrow {\rm Im E} - 0.3$ (this corresponds to dissipators of the form $\sqrt{\kappa} \hat{a}_j$, $\kappa = 0.5$). 
    Plotted are the diagonal covariance matrix elements $\langle \Delta \hat q_j^2\rangle$ and $\langle \Delta \hat p_j^2\rangle$ for an OBC chain. The $q$ quadrature covariance is exponentially amplified toward the right boundary, while the $p$ quadrature remains uniform across the chain.}
\label{fig:delta2_steady_state}
\end{minipage}

\end{figure*}

Next, we consider the full spectrum of Eq.~\eqref{eq:H_Delta2} for larger values of $\Delta_2$, for both OBC and PBC.  We plot two examples in Fig.~\ref{fig:spectrum_D2}. We fix $w=1, \Delta = 0.25, L = 80$, and plot the spectrum for $\Delta_2 = 0.2$ in Fig.~\ref{fig:spectrum_D2}a, and $\Delta_2 = 0.4$ in Fig.~\ref{fig:spectrum_D2}b.
As we have complete decoupling of $q$ and $p$ quadratures, each complex band can be associated with a single quadrature. 
The $q$ band is plotted in olive, with the PBC (OBC) spectrum being the translucent (opaque) curve. The $p$ band is in magenta, with the PBC (OBC) spectrum being the translucent (opaque) curve. We see that the model clearly exhibits the skin effect for both values of $\Delta_2$, in the sense that the spectra exhibit a drastic difference for different boundary conditions. 

Despite still having quadrature decoupling like the bare-BKC, there is a striking difference in the plotted spectra of the $\Delta_2$-BKC: the $q$ and $p$ bands {\it do not} perfectly overlap here, as they would for the bare BKC.  We note that this is generically the case when when adding terms that couple across an even number of sites, even while maintaining quadrature decoupled models. Mechanistically, this comes from the fact that in the equations of motion for quadratures even distance coupling terms contribute with the same sign to both $q$ and $p$ equations, whereas odd distance coupling terms contribute with an opposite sign for the $q$ equation versus the $p$ equation.  As a result, having both kinds of terms breaks the symmetry between $q, p$ equations of motion.

We now provide some comments on winding numbers. This model illustrates why it is important to consider two winding numbers in Eq.~\ref{eq:doubled_winding}. Let us consider the spectra in Fig.~\ref{fig:spectrum_D2}a, and pick $E_{\star}$ to be the eigenvalue of the $q$-band in OBC with maximal imaginary part (in other words, the peak on the OBC curve). One may (visually or otherwise) verify that while this mode is localized, we have that $\nu_q(E_{\star}) = 1$ but $\nu_p(E_{\star}) = 0$. Second, Fig.~\ref{fig:spectrum_D2}b illustrates the importance of recognizing which symmetry enables the skin effect. In particular, we may have used qPHS to write down a winding number, say $\nu_{\rm qPHS}$. Now, qPHS singles out $E = 0$ as a special point, and the model depicted in Fig.~\ref{fig:spectrum_D2}b thus has $\nu_{\rm qPHS} = 0$, despite all the OBC eigenstates being localized, indicating the irrelevance of $\nu_{\rm qPHS}$ in this problem. In these models, the PBC version of each band winds around the entirety of the OBC eigenvalues associated with the same quadrature but not the other; this should be contrasted with the spectra of the models in Fig.~\ref{fig:g3-spectrum}.

Finally, we point out one physically interesting feature of this model -- the maximum imaginary eigenvalue in PBC bands are different for the $q$ and the $p$ bands. To see that this can give rise to interesting phase selected phenomena, we briefly consider the dissipative setting, where we introduce uniform loss to each site. This leads to the effective dynamical matrix
\begin{equation}\label{eq:M_eff}
    M_{\rm eff} = M - i \kappa I,
\end{equation}
from which one may obtain steady state quantities (when the system is stable) by solving a Lyapunov equation, see eg. Ref.~\cite{McDonald_PRB_2022_ness, Lee_PRB_2023_anomalous_relaxation} for more details. From $M_{\rm eff}$, we see that the spectrum of the lossy system can be obtained by translating the spectrum of $M$ along the imaginary axis by $- i \kappa$. In the dissipative version of the system, it is the winding around $0$ (for each decoupled band) that determines whether the system experiences directional amplification \cite{Wanjura_PRL_2021_nonhermitian_dissipative_winding, Wanjura_NatComm_2020_topological_framework_directional_amplification, Porras_PRL_2019_topological_amplification, Ramos_PRA_2021_inputoutput_directional_amplification}. Due to the asymmetry between the two quadratures, one is able to selectively amplify states for the OBC system in the $q$ band without amplifying the $p$ band. This is done by carefully selecting the loss so that only the PBC band for the $q$ quadrature crosses the $E = 0$ line (all other PBC and OBC bands should lie below it). This phenomena is demonstrated in Fig.~\ref{fig:delta2_steady_state}, where the diagonal part of the steady state covariance matrix in the presence of background loss is plotted, with parameters chosen to satisfy the above conditions. For details on the NHSE in the presence of dissipation as well as obtaining such steady state quantities, we refer the reader to eg. Refs.~\cite{Wanjura_2023_quadrature_nonreciprocity, Wanjura_PRL_2021_nonhermitian_dissipative_winding, Lee_PRB_2023_anomalous_relaxation}. Indeed we see that the $q$ quadrature strongly exhibits localization and amplification, whereas the $p$ quadrature remains fairly featureless. Hence, the phase-sensitive winding number translates in this case into a phase-sensitive directional amplification. It would be interesting to study such systems in the presence of nonlinearity as a possible route to phase-sensitive condensate physics \cite{Belyansky_PRL_2025_phase_transitions_nonreciprocal}. 

\subsection{Sublattice symmetry-broken and quadrature-coupled
\label{subsec:with_g0}}
\label{ssec:both_sym_broken}

Finally, we consider the last row of Table~\ref{table:symmetric_schematic}, where both the qPHS and TRS are broken. The simplest way to realize this situation is by introducing a uniform on-site chemical potential (or local detuning). From the general arguments presented above, we therefore expect both the stability protection and the NHSE to be lost. This scenario is also the most relevant experimentally, since small on-site detunings are unavoidable in realistic implementations. It is therefore instructive to analyze in detail how the instability develops. In particular, we will see that in that case, instability can emerge under OBC even when the corresponding PBC system remains completely stable.  This is in sharp contrast to the usual phenomenology of the NHSE. 

Our starting point is the BKC Hamiltonian supplemented by a uniform chemical potential,
\begin{align}\label{eq:H.chem}
\hat{H}_{g_0}
=\hat{H}_{{\rm BKC}}
+g_0\sum_{j=1}^{N}\hat{a}_{j}^{\dagger}\hat{a}_{j}.
\end{align}
Since the bare BKC with OBC is already unstable for $\Delta>w$, we restrict our analysis to the regime $\Delta<w$, where the effect of the chemical potential can be isolated.

\paragraph{Periodic boundary conditions.}
For PBC, the Hamiltonian is diagonal in momentum space after a Fourier transform. The corresponding eigenvalues are
\begin{equation}
\varepsilon_{k}^{\pm}
=
-w\sin k
\pm
i\sqrt{\Delta^{2}\cos^{2}k-g_0^{2}},
\end{equation}
with $k=2\pi n/N$ and $n\in\{1,\ldots,N\}$.
A mode with momentum $k$ becomes unstable whenever
$\Delta|\cos k|>g_0$ (assuming $\Delta,g_0\ge0$). The first instability occurs at $k=0$, yielding the stability criterion
\[
g_0>\Delta.
\]

\paragraph{Open boundary conditions.}

Under OBC, translational invariance is lost and the system cannot be diagonalized by a Fourier transform. We therefore apply the perturbative approach developed in Sec.~\ref{subsec:perturbation in pairing}. The dynamical matrix in the particle-hole basis is
\begin{align}
M
:=
\frac12
\begin{pmatrix}
m & V\\
V & -m^T
\end{pmatrix},
\end{align}
where $m$ and $V$ are $N\times N$ matrices given by
\[
m
=
2g_0\mathbb{I}
+
\sum_j
\left(iwe_j+{\rm H.c.}\right),
\qquad
V
=
i\Delta
\sum_j
\left(e_j+e_j^T\right),
\]
with $e_j=|j\rangle\langle j+1|$.
We further define the bare matrix and the perturbation
\[
M^{(0)}
=
M(\Delta=0)
=
m\oplus(-m^T),
\qquad
M^{(1)}
=
\sigma_x\otimes V.
\]

The eigenstates of the particle-conserving dynamics $m$ (again for OBC) are given by standing waves:

\begin{align}
\varepsilon_k
&=
g_0-w\cos k,
\\
|\varepsilon_k\rangle
&=
\sqrt{\frac{2}{N+1}}
\sum_{j=1}^{N}
i^{-j}
\sin(kj)
|j\rangle,
\end{align}
where $k=\pi n/(N+1)$ and $n\in\{1,\ldots,N\}$.
The corresponding eigenvectors of $M^{(0)}$ are denoted
$|\varepsilon_k,\bullet\rangle
=
|\varepsilon_k\rangle\oplus|0\rangle$
and
$|-\varepsilon_k,\circ\rangle
=
|0\rangle\oplus|\overline{\varepsilon_k}\rangle$.

We now consider the form of the pairing terms $V$ in this standing wave basis.  We find:
\begin{equation}
V
=
\frac{\Delta}{2}
\sum_{k,k'}
f(k,k')
|k,\bullet\rangle
\langle k',\circ|^{*},
\end{equation}
where
\begin{equation}
f(k,k')
=
\frac{1+(-1)^{n-n'+N}}{N+1}
\,
\frac{\sin^2k+\sin^2k'}{\cos k+\cos k'},
\label{eq:expressiong}
\end{equation}
for $k\neq\pi-k'$, while $f(k,\pi-k)=0$.

As discussed in Sec.~\ref{subsec:perturbation in pairing}, the pairing matrix $V$ only couples particle and hole sectors. Consequently, all non-degenerate eigenvalues receive vanishing first-order corrections in $\Delta$. A degenerate subspace is obtained whenever a particle with energy $\varepsilon_k$ is resonant with a hole of energy $-\varepsilon_{k'}=\varepsilon_k$, which requires
\begin{equation}
\cos k+\cos k'
=
\frac{2g_0}{w}.
\label{eq:selectionrule}
\end{equation}
Because the chemical potential breaks the sublattice symmetry, these resonant particle-hole pairs generally become unstable. The first-order correction is
\begin{equation}
\delta\varepsilon_k
=
\pm
i\frac{\Delta}{2}
f(k,k'),
\label{eq:delteps}
\end{equation}
which is purely imaginary whenever $f(k,k')\neq0$.

For fixed $g_0$ and $k'$, Eq.~\eqref{eq:selectionrule} uniquely determines $k$. In the thermodynamic limit, unstable modes satisfy
\[
k\in
\left[
0,
\cos^{-1}\!\left(\frac{2g_0}{w}-1\right)
\right],
\]
so the unstable momentum window continuously shrinks as the chemical potential increases.

\begin{figure}[hptb]
\centering
\includegraphics[width=0.98\columnwidth]{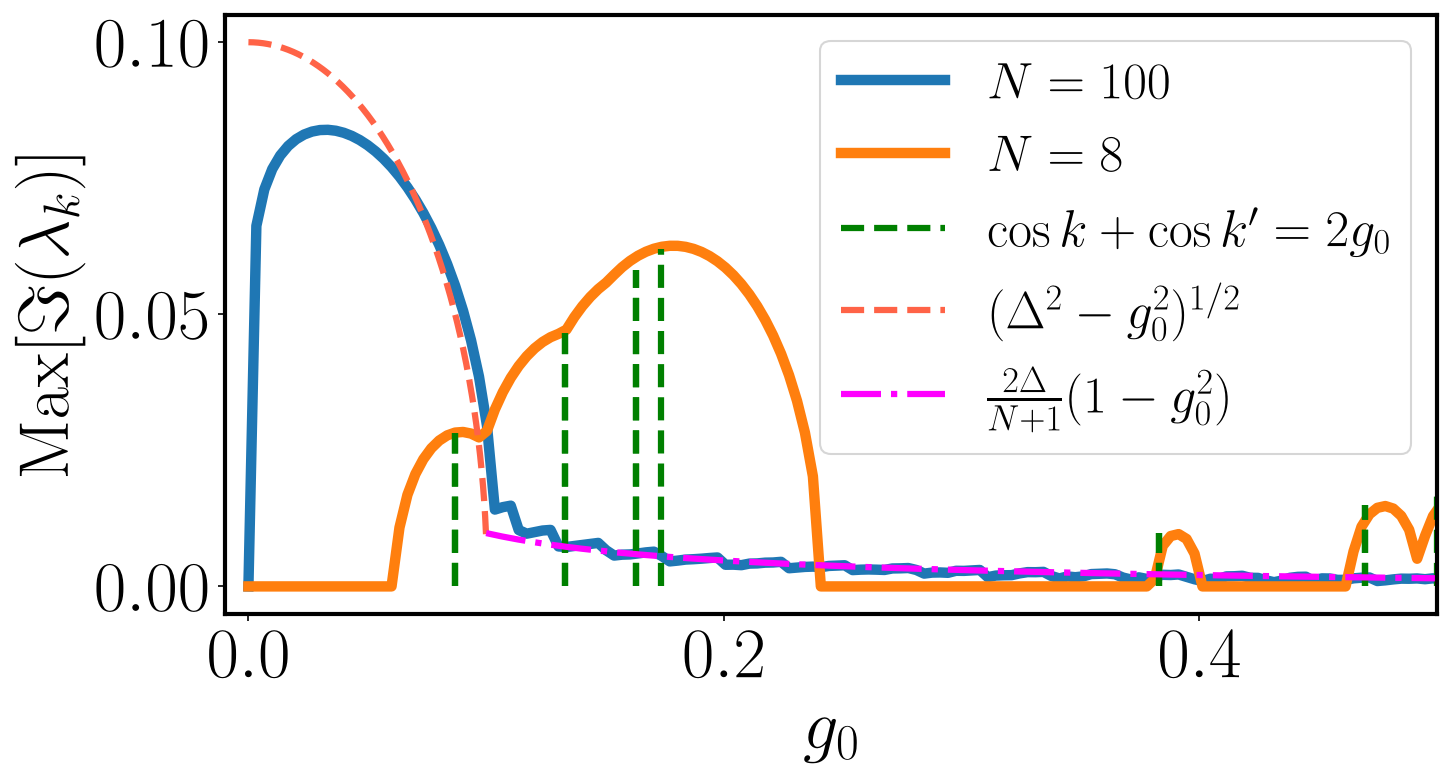}
\caption{Plot of maximum imaginary part of the eigenvalues $\lambda_k$ for the $g_0$-BKC (see Eq.~\eqref{eq:H.chem}) as a function of $g_0$ for $N=100$ (blue) and $N=8$ (orange), with $\Delta=0.1$ and $w=1$. The dashed green curve indicates the resonance condition $\cos k+\cos k'=2g_0/w$, while the dashed red curves show the coarse-grained large-$N$ prediction.}
\label{fig:Maximum-of-theRealpartofeigenvalue}
\end{figure}

Finite-size effects are particularly pronounced for small $N$. Since $g_0$ is a continuous parameter whereas the allowed momenta are discrete, the resonance condition may fail to be satisfied, allowing the system to alternate between stable and unstable regions as $g_0$ is varied. This behavior is illustrated in Fig.~\ref{fig:Maximum-of-theRealpartofeigenvalue} for $N=8$\footnote{For $g_0=0$, the resonance condition reduces to $\cos k+\cos k'=0$, corresponding to opposite momenta $k$ and $\pi-k$, i.e., $n'=N+1-n$. Since $f(k,\pi-k)=0$, these resonant pairs never produce an instability at first order, consistent with the known behavior of the bare BKC.}.

For sufficiently large systems and $g_0<w$, there is always at least one unstable mode. Moreover, in the large-$N$ limit the spectrum can be computed quantitatively beyond perturbation theory, even at finite $\Delta$, by means of a coarse-graining approach described in App.~\ref{app_sec:large_N_chem}. The resulting expression for the largest instability is
\begin{align}
\max\!\left[\Im(\lambda_k)\right]
=
\begin{cases}
\sqrt{\Delta^2-g_0^2},
&
g_0<\Delta,
\\[1ex]
\dfrac{2\Delta}{N+1}\left(1-g_0^2\right),
&
g_0>\Delta.
\end{cases}
\end{align}
Figure~\ref{fig:Maximum-of-theRealpartofeigenvalue} shows excellent agreement between this prediction and the exact numerical spectrum for $N=100$.

These results reveal that, in the thermodynamic limit, the OBC instability profile approaches that of the PBC system but acquires an additional tail of weak instabilities whose amplitude scales as $N^{-1}$. Remarkably, throughout the regime $g_0>\Delta$, the PBC system is completely stable whereas the OBC system remains weakly unstable which is the opposite of the conventional NHSE scenario.

\section{Conclusion}\label{sec:conclusion}

We have shown that the striking non-Hermitian phenomenology of the bosonic Kitaev chain (BKC) (including its non-Hermitian skin effect and its anomalous sensitivity to boundary conditions) can be traced to two independent symmetries. The first, an effective particle-hole symmetry (qPHS), is equivalent to a decoupling of the dynamics of canonical quadratures and reduces the problem to two copies of the Hatano-Nelson model. The second more general mechanism is an effective time-reversal symmetry that follows {\it automatically} from the sublattice (chiral) symmetry of any quadratic bosonic pairing Hamiltonian.  This symmetry protects a genuine symmetry-protected skin effect that in general is not reducible to single-band physics. We further showed that this same sublattice symmetry perturbatively protects the BKC against dynamical instability under open boundary conditions, giving a unifying symmetry principle behind the model's coexisting skin effect and stability.

Because quadrature decoupling and sublattice symmetry coincide only accidentally in the bare BKC, treating them as independent design principles allowed us to construct and classify a broader family of BKC-like models. These exhibit a richer range of phenomena than the original model, including localization transitions with no counterpart in the bare BKC, phase-sensitive and frequency-selective amplification, and instabilities whose open and periodic boundary phase diagrams can disagree in counterintuitive ways. We also established a precise correspondence between a hopping-augmented BKC and the symplectic Hatano-Nelson model, showing that a symmetry with no physical time-reversal interpretation in the bosonic setting corresponds to a bona fide time-reversal symmetry in an associated fermionic model.

Several directions follow naturally from this work. On the formal side, it would be worth clarifying the relationship between the spectral winding numbers used here (which are natural to the non-Hermitian setting) and the wavefunction-based winding numbers of Hermitian topology, particularly in regimes where dynamical instability is present. Extending the symmetry classification to interacting models, beyond the mean-field treatment sketched in App.~\ref{sec:interactions} is another natural next step, as is a non-perturbative treatment of the finite-range OBC stability observed numerically in the $g_3$-BKC model. Finally, given the connections to parametric amplification and sensing noted throughout, it would be interesting to explore whether the frequency-selective and phase-sensitive instabilities of these generalized models offer practical advantages over the bare BKC in such applications.

\section*{Acknowledgements}

This work was supported by the Air Force Office of Scientific Research under Grant No. FA9550-19-1-0362, and by the Army Research Office under grant W911NF-25-1-0286. A.C. also acknowledges support from the Simons Foundation through a Simons Investigator Award (Grant No. 669487, A.C.). TJ is funded by the ANR-25-CE57-2088 JCJC (QuDi).


\bibliography{references}

\appendix



\section{Large $N$ coarse-graining for the model presented in Sec.~\ref{subsec:with_g0}}\label{app_sec:large_N_chem}

Beyond perturbative method, it is possible to compute quantitatively
the spectrum at finite $\Delta$ in the large system size limit by
use of a coarse-graining approach.

The spirit of the method is to identify the pairs of momenta such
that the real part of the spectrum is the largest, to expand $M$
in powers of $\frac{1}{N}$ over these values and to diagonalize the
main contribution. Intuitively, we expect that the imaginary part
of the spectrum will take large values when either of the two conditions
are fulfilled, or close to being fulfilled: 
\begin{align}
({\rm I}):\cos k+\cos k' & =0,\\
({\rm II}):\cos k+\cos k' & =\frac{2g_0}{w},
\end{align}
(I) corresponds to the pairs for which $f(k,k')$ is largest and (II)
to the degeneracy condition.

The ``block structure'' coupling the modes $\left|k,\bullet\right\rangle $
and $\left|k',\circ\right\rangle $ is of the following form 
\begin{equation}
m_{k,k'}:=\begin{pmatrix}g_0-t\cos k  & \frac{\Delta}{2}g(k,k')\\
-\frac{\Delta}{2}f(k,k') & g_0-t\cos k'
\end{pmatrix}.
\end{equation}
We begin by examining condition (I). Let us expand $k'=\pi-k+\varepsilon$
where $\varepsilon:=\frac{\pi m}{N+1}$ with $m\in\mathbb{N}$ and
$m\ll N$. Then, to leading order in $N^{-1}$: 
\begin{equation}
f(k,k')\approx\frac{2}{\pi m}\sin k\left(1+(-1)^{m+1}\right).\label{eq:coarsegrained}
\end{equation}
We can interpret $k$ as a ``slow variable'' and $m$ as a fast
variable in the expression above.

As we now show, an $M\times M$ matrix $O$ with elements
\begin{equation}
O_{k,k'}:=i\frac{1+(-1)^{k-\left(M+1-k'\right)}}{k-(M+1-k')}
\end{equation}
can be rotated, up to finite size correction, to
\begin{equation}
O\approx UXU^{-1}
\end{equation}
with $U$ an invertible matrix and $X$ the constant $M\times M$
antidiagonal matrix with elements 
\begin{equation}
X_{k,k'}=\delta_{k,M+1-k'}.
\end{equation}
For simplicity, we suppose $M$ to be even. Consider then the $Z$
diagonal matrix with $1$ up to $M/2$ and $-1$ from $M/2+1$ to
$M$: 
\begin{align}
Z & :=\sum_{j=1}^{M}\operatorname{sgn}(M+1-2j)\left|j\right\rangle \left\langle j\right|e^{i\pi j},\nonumber\\
 & =\frac{1}{M}\sum_{k,k',j=1}^{M}\operatorname{sgn}(M+1-2j)\left|k\right\rangle \left\langle k'\right|e^{-i\frac{2\pi j}{M}(k-k')},\nonumber\\
 & \approx i\sum_{k,k'=1}^{M}\frac{\left(1+(-1)^{k-k'+1}\right)}{\pi(k-k')}\left|k\right\rangle \left\langle k'\right|.
\end{align}
where in the second line we introduced the PBC Fourier transform
$\left|j\right\rangle :=\frac{1}{\sqrt{M}}\sum_{k=1}^{M}e^{-i\frac{2\pi jk}{M}}\left|k\right\rangle $
and in the third line we took the continuous limit to turn the sum over $j$ into an integral. 

We thus see that
the subblock of $V$ centered around $k$ can be rotated
to a constant antidiagonal matrix where the constant is $i\Delta\sin k$.
Thus, the total matrix $M$ admits in this rotated basis the expression
\begin{equation}
M\approx\oplus_{k}\begin{pmatrix}g_0-t\cos k  & -\Delta\sin k\\
\Delta\sin k & - g_0-t\cos k 
\end{pmatrix}.
\end{equation}
The eigenvalues of these blocks are given by: 
\begin{equation}
\varepsilon_{k}^{\pm}=-t\cos k\mp i\sqrt{\Delta^{2}\sin^{2}k-g_0^{2}}
\end{equation}
so we restore the same stability criterion as for PBC, that is $g_0<\Delta$.
Remark that this approximate approach misses the fact the system \emph{is}
stable for $g_0=0$.

For the condition (II), let $k^{*}:=\cos^{-1}\left(\frac{2g_0}{w}-1\right)$,
and $k^{*}=:\frac{\pi m^{*}}{N+1}$, $m^{*}\in\mathbb{N}$ and expand
$k'=k^{*}+\varepsilon$. To leading order in $N^{-1}$: 
\begin{equation}
f(k,k')\approx\frac{\left(1+(-1)^{-m^{*}-m+N}\right)}{N+1}h\left(k,\frac{2g_0}{w}\right),\label{eq:coarsegrained-1}
\end{equation}
where we abbreviated $h(p,x):=\frac{2\sin^{2}p-x^{2}+2x\cos p}{x}$
and the corresponding eigenvalues of the block centered around $\left(k,k^{*}\right)$
are 
\begin{equation}
\varepsilon_{k}^{\pm}=g_0-t\cos k\pm i\frac{\Delta\left(1+(-1)^{-m^{*}-m+N}\right)}{2(N+1)}f\left(k,\frac{2g_0}{w}\right)
\end{equation}
The most unstable mode is attained for $\cos k=\frac{g_0}{w}$ for
which $\Im\left(\varepsilon_{k}^{+}\right)=\frac{2\Delta}{N+1}\left(1-\left(\frac{g_0}{w}\right)^{2}\right)$.
We thus see that condition (II) will generically lead to instabilities
as long as $g_0<w$. 

To summarize, the OBC system is \emph{always unstable} for $g_0<w$.
However, the unstable modes can come either from the resonance conditions
(I) or (II). Condition (I) produces an instability of order $0$ in
$N^{-1}$ but only for $g_0<\Delta$. The condition (II) can always
be fulfilled for large system sizes but yields a weaker instability
of order $1/N$. We checked that our analysis is consistent with numerical
diagonalization of the problem, see Fig.~\ref{fig:Maximum-of-theRealpartofeigenvalue}.

\section{Comments on interacting models}\label{sec:interactions}

We note that one key difference between sublattice symmetry and quadrature decoupling is that sublattice symmetry can also be used to study some interacting models. For instance, one basic set of interactions that can be studied in this framework are the density dependent versions of various quadratic terms. In particular, consider interacting versions of each term in Eq.~\eqref{eq:H.BKC.g1}: (1) A density-dependent chemical potential (Hubbard-U), which breaks sublattice symmetry, (2) Density-dependent (a) imaginary hopping or (b) real hopping terms, which preserve sublattice symmetry. This involves adding the following terms to Eq.~\eqref{eq:H.BKC.g1}.
\begin{equation}
    \begin{aligned}
        (1): &U \sum_j \hat{a}_j^{\dag} \hat{a}_j \hat{a}_j^{\dag} \hat{a}_j, \\
        (2a): &\frac{i w_{\rm int}}{2} \sum_j \left( \hat{a}_j^{\dag} \hat{a}_j + \hat{a}_{j+1}^{\dag} \hat{a}_{j+1} \right) \hat{a}^{\dag}_{j+1} \hat{a}_j + {\rm H.c.}, \\
        (2b): &\frac{g_{\rm int}}{2} \sum_j \frac{1}{2} \left( \hat{a}_j^{\dag} \hat{a}_j + \hat{a}_{j+1}^{\dag} \hat{a}_{j+1} \right) \hat{a}^{\dag}_{j+1} \hat{a}_j + {\rm H.c.}.
    \end{aligned}
\end{equation}

On the second-quantized level, it is meaningful to say that the term (1) breaks sublattice symmetry, whereas (2a), (2b) do not. Furthermore, on the mean-field level, the equations of motion also retain the character of the sublattice symmetry -- adding a $(-1)^j$ phase flips their signs in the same way as the quadratic analogues. Conversely, the mean-field equations of motion of any quartic term necessarily couples quadratures, so it is in general not possible to extend the single-band analysis via quadrature decoupling into the interacting regime. We leave more detailed study of these models to future work.

\end{document}